\documentclass[aps,nofootinbib,floatfix,showpacs,preprintnumbers,twocolumn]{revtex4} 
\usepackage{graphicx}
\usepackage{bm}
\usepackage{mathrsfs,amssymb}
\usepackage{float}
\usepackage{tabularx} 
\usepackage{amsmath}
\usepackage{epstopdf}
\usepackage{multirow}
\usepackage{color}
\usepackage{booktabs}
\usepackage{hyperref}
\usepackage{mathtools}
\DeclarePairedDelimiter\floor{\lfloor}{\rfloor}
\usepackage{tabularx} 
\input epsf

\def\lsim{\mathrel{\raise.3ex\hbox{$<$\kern-.75em\lower1ex\hbox{$\sim$}}}}
\def\gsim{\mathrel{\raise.3ex\hbox{$>$\kern-.75em\lower1ex\hbox{$\sim$}}}}

\def\cmm2{{\,\rm cm^{-2}}}
\def\cm2{{\,{\rm cm}^2}}
\def\cmm3{{\,{\rm cm}^{-3}}}
\def\gcmm3{{\,{\rm g\,cm^{-3}}}}

\def\fun#1#2{\lower3.6pt\vbox{\baselineskip0pt\lineskip.9pt
  \ialign{$\mathsurround=0pt#1\hfil##\hfil$\crcr#2\crcr\sim\crcr}}}

\def\etal{{\it et al.}}

\def\be{\begin{equation}}
\def\ee{\end{equation}}
\def\bea{\begin{eqnarray}}
\def\eea{\end{eqnarray}}

\begin{document}

\title{Axion astronomy with microwave cavity experiments}

\author{Ciaran A. J. O'Hare}\email{ciaran.ohare@nottingham.ac.uk} \affiliation{School of Physics and Astronomy, University of Nottingham, University Park, Nottingham, NG7 2RD, United Kingdom}

\author{Anne M. Green}\affiliation{School of Physics and Astronomy, University of Nottingham, University Park, Nottingham, NG7 2RD, United Kingdom}

\date{\today}
\smallskip
\begin{abstract}
Terrestrial searches for the conversion of dark matter axions or axion-like particles into photons inside magnetic fields are sensitive to the phase space structure of the local Milky Way halo. We simulate signals in a hypothetical future experiment based on the Axion Dark Matter eXperiment (ADMX) that could be performed once the axion has been detected and a frequency range containing the axion mass has been identified. We develop a statistical analysis to extract astrophysical parameters, such as the halo velocity dispersion and laboratory velocity, from such data and find that with only a few days integration time a level of precision can be reached matching that of astronomical observations. For longer experiments lasting up to a year in duration we find that exploiting the modulation of the power spectrum in time allows accurate measurements of the Solar peculiar velocity with an accuracy that would improve upon astronomical observations. We also simulate signals based on results from N-body simulations and find that finer substructure in the form of tidal streams would show up prominently in future data, even if only a subdominant contribution to the local dark matter distribution. In these cases it would be possible to reconstruct all the properties of a dark matter stream using the time and frequency dependence of the signal. Finally we consider the detection prospects for a network of streams from tidally disrupted axion miniclusters. These features appear much more prominently in the resolved spectrum than suggested by calculations based on a scan over a range of resonant frequencies, making the detection of axion minicluster streams more viable than previously thought. These results confirm that haloscope experiments in a post-discovery era are able to perform ``axion astronomy''.
\end{abstract}
\pacs{14.80.Va; 95.35.+d}
\maketitle

\section{Introduction}
Axions are light pseudoscalar particles that appear in the solution of Peccei and Quinn~\cite{Peccei:1977hh, Kim:2008hd} to explain the unnatural absence of CP-violation in quantum chromodynamics (QCD). More modern motivation from the landscape of axion-like particles (ALPs) appearing in string theory~\cite{Svrcek:2006yi,Arvanitaki:2009fg,Cicoli:2012sz}, inspires the generalisation of axions to light pseudoscalars with a theoretical origin outside of the original Peccei-Quinn solution. Such ALPs can cover an extremely wide range of masses and couplings to the standard model~\cite{Arias:2012az}. Axions and ALPs are an attractive cold dark matter candidate and can be produced in the early Universe through a variety of non-thermal mechanisms such as vacuum misalignment or the decay of topological defects~\cite{Davis:1986xc,Hiramatsu:2012gg} in ways that are consistent with the known cosmological abundance and phenomenology of dark matter~\cite{Ipser:1983mw,Wantz:2009it}. For a recent review of axion cosmology see e.g., Ref.~\cite{Marsh:2015xka}. In the following we use the term `axion' to refer to both the QCD axion and generic axion-like particles.

Certain axion mass and coupling ranges can be ruled out with various astrophysical observations such as the cooling of white dwarfs~\cite{Raffelt:1985nj,Isern:2008nt}, neutron star interactions~\cite{Berenji:2016jji}, the lifetimes of horizontal branch stars in globular clusters~\cite{Raffelt:1999tx} and supernovae neutrinos~\cite{Burrows:1988ah,Mayle:1987as}. Axions may also be observable in the lab. Experimental tests for axions predominantly rely on their coupling to electromagnetism resulting in the Primakoff conversion of axions into photons inside strong magnetic fields. Cavity resonators can exploit this effect if the resonant frequency is chosen to match the axion mass~\cite{Sikivie:1983ip}. As the axion mass is unknown experiments must be designed to scan over a range of resonant frequencies corresponding to a range of axion masses. Experiments include the helioscope CAST~\cite{Zioutas:2004hi} (and the planned IAXO~\cite{Armengaud:2014gea}) searching for axions produced inside the Sun, and haloscopes such as ADMX searching for Galactic dark matter axions~\cite{Asztalos:2009yp}. These experiments operate over a narrow range of frequencies and hence make constraints on the axion mass in small bands, where the smallest accessible photon coupling is controlled by the signal-to-noise level of the experiment. Planned dark matter axion experiments such as QUAX~\cite{Barbieri:2016vwg,Crescini:2016lwj,Ruoso:2015ytk}, CULTASK~\cite{Chung:2016ysi} and the layered dielectric haloscope MADMAX~\cite{TheMADMAXWorkingGroup:2016hpc,Millar:2016cjp} are being designed to probe ranges of axion masses inaccessible due to the technical restrictions of the ADMX design. As well as axion `observation' experiments there exist solely lab-based experiments such as the Any Light Particle Search~\cite{Bahre:2013ywa} using the technique known as light-shining-through-a-wall~\cite{VanBibber:1987rq}. In haloscopes beyond ADMX, it may be possible to search for lighter axions with broadband readout circuits~\cite{Kahn:2016aff} or LC circuits~\cite{Sikivie:2013laa} as well as heavier masses in the meV range with Josephson junctions~\cite{Beck:2014aqa,Beck:2013jha}. Other couplings such as those to nuclei~\cite{Arvanitaki:2014dfa} can be probed in nuclear magnetic resonance experiments such as CASPEr~\cite{Budker:2013hfa,Graham:2013gfa} and the coupling to electrons can be constrained using conventional dark matter direct detection searches for weakly interacting massive particles (WIMPs)~\cite{Aprile:2014eoa}. For a recent review of axion experiments see for example Ref.~\cite{Graham:2015ouw}.

The goal of a haloscope experiment is to tune the frequency of a cavity mode to the axion mass resulting in the resonant enhancement of the axion-photon conversion. ADMX achieves this with the use of movable tuning rods placed inside the cavity itself to modulate the resonant frequency over a range of several hundred MHz. Although usually unimportant when scanning over a relatively large range of resonant frequencies, the velocity distribution of axions in the halo would cause a small frequency spread in the resonance~\cite{Krauss:1985ub}. Furthermore there are also 0.5\% and 1\% modulations in time due to the relative velocity of the Earth and Sun with respect to the halo dark matter `wind'~\cite{Turner:1990qx,Ling:2004aj,Vergados:2016rlh}. There is also a potentially exploitable $\mathcal{O}(1)$ modulation dependent on cavity orientation with respect to the incoming axion direction for axion masses (and haloscope volumes) experiencing a loss of coherence over the cavity dimensions~\cite{Irastorza:2012jq}. 

Here we consider a scenario in which the axion mass has been determined down to a level of precision dictated by the scanning approach of ADMX and a dedicated high spectral resolution experiment is then performed at a single resonant frequency. In such a situation, the shape of the spectrum of axion-photon conversion will be accessible. This power spectrum is related to the velocity distribution of the local dark matter halo, hence the precise functional form and parameters which arise from astrophysics are important. Past axion searches with ADMX have incorporated some of these astrophysical uncertainties, for example by searching for discrete flows of axions~\cite{Duffy:2006aa,Hoskins:2016svf,Hoskins:2011iv} or applying constraints to different halo models~\cite{Sloan:2016aub,Vergados:2016rlh}. In the situation we consider here however it will be possible to perform ``axion astronomy'' in the sense that a measurement can be made directly of the axion power spectrum to learn about the structure of dark matter in the Galaxy. For this reason we develop an analysis that shares similarity with well-established methods for extracting astrophysical information in the case of WIMP direct detection experiments e.g., Ref.~\cite{Strigari:2009zb}. Since axion haloscopes effectively observe the axions directly - as opposed to WIMP direct detection experiments which observe the WIMP flux convolved with a stochastic scattering process - the prospects for sensitive measurements of the dark matter halo are much greater. Here, we show how powerful future ADMX-like experiments might be for doing axion astronomy. We discuss this idea in the context of simple analytic halo models, distributions from N-body simulations, and minicluster streams - a phenomenon unique to axion dark matter~\cite{Hogan:1988mp}.

To begin in Sec.~\ref{sec:theory} we review some of the basic theory for axions and ALPs, as well as the laboratory frame speed distribution relevant for calculating signals inside a haloscope experiment. In Sec.~\ref{sec:experiment} we outline the steps in calculating the expected power inside a magnetic cavity resonating at a given frequency. Then in Sec.~\ref{sec:analysis} we describe our mock experiment and explain the procedure used to extract astrophysical information from the simulated data. The first results applying these methods to the reconstruction basic sets of input parameters are presented in Sec.~\ref{sec:reconstruction} and then extended to N-body data from the Via Lactea II (VL2)~\cite{Diemand:2007qr} simulation in Sec.~\ref{sec:nbody}. Finally we extend this simulation to tidal streams from disrupted axion miniclusters in Sec.~\ref{sec:miniclusters}, before summarising in Sec.~\ref{sec:summary}.

\section{Axions and ALPs}\label{sec:theory}
First we outline some of the essential steps in calculating the resonantly enhanced axion-photon conversion power inside a microwave cavity. Full details of these calculations can be found in Refs.~\cite{Hong:2014vua,McAllister:2015zcz,Krauss:1985ub}. Importantly we wish to make the connection to realistic halo velocity distributions so we depart from an often used approximation that the axion power spectrum can be described with a Breit-Wigner function.

The axion to two photon coupling is given by the formula,
\begin{equation}
 g_{a\gamma\gamma} = \frac{g_\gamma \alpha}{\pi f_a} \, ,
\end{equation}
which includes the fine structure constant $\alpha$ and the axion decay constant $f_a$. The dimensionless coupling $g_\gamma$ is,
\begin{equation}
 g_\gamma = \frac{1}{2}\left(\frac{E}{N} - \frac{2}{3}\frac{4+z}{1+z}\right) \, .
\end{equation}
In which  $E/N$ is the ratio of the colour axion anomaly to the electromagnetic axion anomaly and $z$ is the ratio of the up and down quark masses. The value of this constant is model dependent: $E/N=-0.97$ for the KSVZ model~\cite{Kim:1979if,Shifman:1979nz} and $0.36$ for the DFSZ model~\cite{Dine:1981rt,Zhitnitsky:1980tq} for example. In the interest of model independence and to generalise to ALPs we express the interaction in terms of the coupling $g_{a\gamma\gamma}$.

The effective Lagrangian for axions coupled to electromagnetism is,
\begin{eqnarray}
 \mathcal{L} =& \frac{1}{2}\partial_\mu a \partial^\mu a - V(a) + \frac{1}{4}g_{a\gamma\gamma} a F_{\mu\nu}\tilde{F}^{\mu\nu} \nonumber \\
    &- \frac{1}{4}F_{\mu\nu}F^{\mu\nu} + j^\mu A_\mu + a\rho_q \, ,
\end{eqnarray}
where $F_{\mu\nu}$ is the electromagnetic field strength tensor, and $\tilde{F}^{\mu\nu} = \frac{1}{2}\epsilon^{\mu\nu\rho\sigma}F_{\rho\sigma}$ its dual. The axion potential $V(a)$ is provided by QCD instanton effects and can be approximated with a simple mass term $\frac{1}{2} m_a^2 a^2$. The axionic charge density and the electromagnetic current density are written as $\rho_q$ and $j^\mu$. Writing $F_{\mu\nu}\tilde{F}^{\mu\nu} = -4\,\textbf{E}\cdot\textbf{B}$ we then see the axion-photon interaction in terms of electric and magnetic field strengths,
\begin{equation}
 \mathcal{L}_{a\gamma\gamma} = - g_{a\gamma\gamma}\, a \textbf{E}\cdot\textbf{B} \, .
\end{equation}
This interaction modifies Maxwell's equations to include an additional axion current,
\begin{eqnarray}
&\nabla \cdot \textbf{E} &= \rho_q + g_{a\gamma\gamma} \nabla a \cdot \textbf{B} \,, \\
&\nabla \cdot \textbf{B} &= 0 \,, \\
&\nabla \times \textbf{E} &= -\frac{\partial \textbf{B}}{\partial t}\,,  \\
&\nabla \times \textbf{B} &= \mu_0\textbf{j} + \frac{\partial \textbf{E}}{\partial t} - g_{a\gamma\gamma} \textbf{B}_0\frac{\partial a}{\partial t} - g_{a\gamma\gamma} \nabla a \times \textbf{E} \, .\quad\quad
\end{eqnarray}
However these equations simplify for the setup we consider here. Firstly we assume the axion field has no spatial dependence on laboratory scales ($\nabla a = 0$). We can do this because the size of ADMX is around the 1~meter scale and is well below the de Broglie wavelength of the axion field for the mass ranges we consider ($>$100~m). This allows us to assume that there is no spatial dependence in the axion field over the dimensions of the cavity and hence no additional modulations due to the changing orientation of the cavity with respect to the axion wind. We also assume that there is no axionic charge and no electromagnetic current inside the cavity: $\rho_q = 0$ and $j^\mu = 0$. This results in the following simple set of equations,
\begin{eqnarray}
&\nabla \cdot \textbf{E} &= 0 \,, \\
&\nabla \cdot \textbf{B} &= 0 \,, \\
&\nabla \times \textbf{E} &= -\frac{\partial \textbf{B}}{\partial t} \,, \\
&\nabla \times \textbf{B} &= \frac{\partial \textbf{E}}{\partial t} - g_{a\gamma\gamma} \textbf{B}_0\frac{\partial a}{\partial t} \, .
\end{eqnarray}

Under the above assumptions the equation of motion for the axion field is,
\begin{equation}
 \Box a \simeq \frac{\partial^2 a}{\partial t^2}= -V'(a) - g_{a\gamma\gamma}\textbf{E}\cdot\textbf{B} \, .
\end{equation}

Dark matter axions in the Milky Way undergo essentially no interactions, so in a quadratic potential $V(a)~\simeq~\frac{1}{2} m_a^2 a^2$, the field oscillates coherently at the axion mass $a(t) = a_0 e^{im_a t} \equiv a_0 e^{i\omega t}$. However due to thermalisation in the Milky Way the coherence of the oscillations is spoiled slightly by a dispersion in axion velocities: $ \omega = m_a(1+\frac{1}{2}v^2 + \mathcal{O}(v^4))$. One can account for this by moving to a Fourier description of the field, written as $\mathcal{A}(\omega)$,
\begin{eqnarray}
 a(t) &=& \sqrt{T} \int_{-\infty}^{+\infty} \frac{\textrm{d}\omega}{2\pi} \mathcal{A}(\omega) e^{-i\omega t} \,, \\
 \mathcal{A}(\omega) &=& \frac{1}{\sqrt{T}} \int_{-T/2}^{T/2} \textrm{d}t \,  a(t) e^{i\omega t} \, ,
\end{eqnarray}
where $T$ is some large reference time used to take the averages. The quantity $|\mathcal{A}(\omega)|^2$ is referred to as the axion power spectrum. The rms of the axion field squared is connected to the axion power spectrum by the Parseval relation,
\begin{equation}\label{eq:parseval}
 \langle a^2(t) \rangle = \frac{1}{T} \int_{-T/2}^{T/2} \textrm{d} t\, a^2(t) = \int_{-\infty}^{+\infty} \frac{\textrm{d}\omega}{2\pi} |\mathcal{A}(\omega)|^2 \, .
\end{equation} 

The convention in dark matter detection literature is to use a velocity distribution to describe the kinematics of the local halo. In this context we must blur the distinction between the interpretation of axionic dark matter as a classically oscillating field and as a collection of particles. The velocity distribution is related to the axion power spectrum in the following way. First we write down the distribution of axion velocities $f_\textrm{lab}(\textbf{v})$ in the laboratory frame (i.e., $f_\textrm{lab}(\textbf{v})~=~f_\textrm{gal}(\textbf{v}~+~\textbf{v}_\textrm{lab}))$ by temporarily introducing a number density,
\begin{equation}
 \textrm{d}n = n_0 f_\textrm{lab}(\textbf{v}) \textrm{d}^3 v \, ,
\end{equation}
where $\textrm{d}n$ is the number density of ``particles'' with speeds between $v$ and $v + \textrm{d}v$. The constant $n_0$ is found from integrating $\textrm{d}n$ over all velocities and is used to define the local axion number density $n_0 \equiv \rho_a/m_a$. This allows the connection to a classical field oscillating at $m_a$, which should have $\langle a^2(t)\rangle = n /m_a $, to be made~\cite{Krauss:1985ub}.

An expression for the axion power spectrum $|\mathcal{A}(\omega)|^2$  can be obtained by satisfying Parseval's relation and changing variables from $\omega$ to $v$,
\begin{equation}
|\mathcal{A}(\omega)|^2 = 2\pi \frac{\textrm{d}\langle a^2(t) \rangle}{\textrm{d}v}\frac{\textrm{d}v}{\textrm{d}\omega} \, ,
\end{equation}
we can then substitute for $\textrm{d}\langle a^2(t) \rangle /\textrm{d}v$ using,
\begin{eqnarray}
 \frac{\textrm{d}n}{\textrm{d}v} &=& n_0 \int v^2 f_\textrm{lab}(\textbf{v}) \textrm{d}\Omega \\
  &=& n_0 f_\textrm{lab}(v) \, ,
\end{eqnarray}
where this expression clarifies the distinction between a 3-dimensional velocity distribution $f(\textbf{v})$ and its 1-dimensional speed distribution $f(v)$. Hence, the formula for the axion power spectrum on Earth can be written as,
\begin{equation}\label{eq:Asqlab}
|\mathcal{A}(\omega)|^2= 2\pi \frac{\rho_a}{m_a^2} f_\textrm{lab}(v)\frac{\textrm{d}v}{\textrm{d}\omega} \, .
\end{equation}

The simplest assumption for a dark matter halo is the Standard Halo Model (SHM) which is spherically symmetric with a $1/r^2$ density profile and yields a Maxwell-Boltzmann velocity distribution,
\begin{equation}
 f_\textrm{gal}(\textbf{v}) = \frac{1}{\pi\sqrt{\pi}} \frac{1}{v_0^3} e^{-|\textbf{v}|^2/v_0^2} \, .
\end{equation}
To simplify the following expressions we use the peak velocity $v_0$ for the shape of the distribution, however it can also be written in terms of a velocity dispersion, $v_0^2~=~2 \sigma_v^2$. The speed distribution follows from an integral over all directions,
\begin{equation}
 f_\textrm{gal}(v) = 4\pi\frac{1}{\pi\sqrt{\pi}} \frac{v^2}{v_0^3} e^{-v^2/v_0^2} \, .
\end{equation}
However the power spectrum should reflect the fact that we observe $f_\textrm{lab}$ not $f_\textrm{gal}$ so we need to compute the speed distribution in the laboratory frame. To do this we make the transformation $\textbf{v} \rightarrow \textbf{v} - \textbf{v}_\textrm{lab}(t)$ which yields for the velocity and speed distributions,
\begin{equation}
 f_\textrm{lab}(\textbf{v},t) = \frac{1}{\pi\sqrt{\pi}} \frac{1}{v_0^3} e^{-(\textbf{v}-\textbf{v}_\textrm{lab}(t))^2/v_0^2} \, ,
\end{equation}
and,
\begin{equation}\label{eq:axionspeeddist}
 f_\textrm{lab}(v,t) = \frac{2v e^{-(v^2+v_\textrm{lab}(t)^2)/v_0^2} \sinh{\left(\frac{2v_\textrm{lab}(t)v}{v_0^2}\right)}}{\pi\sqrt{\pi}v_0 v_\textrm{lab}(t)}.
\end{equation}
Since we are now in the moving laboratory frame a time dependence appears in the speed distribution. Finally after changing variables to $\omega = m_a(1 + v^2/2)$ we arrive at the axion power spectrum. The axion power spectrum must be 0 for $\omega<m_a$ which is enforced by requiring that $f(v)$ be real.

For use in later examples we also define the velocity distribution for streams $f^\textrm{str}_\textrm{lab}(\textbf{v})$ which are spatially and kinematically localised substructure components of the dark matter halo. Their velocity distribution can also be described with a Maxwellian,
\begin{equation}\label{eq:streamfv}
 f^\textrm{str}_\textrm{lab}(\textbf{v},t) = \frac{1}{(2\pi\sigma_\textrm{str}^2)^{3/2}} e^{-(\textbf{v}-(\textbf{v}_\textrm{lab} - \textbf{v}_\textrm{str}(t))^2/2\sigma_\textrm{str}^2} \, ,
\end{equation}
where the stream is parameterised by its Galactic frame velocity $\textbf{v}_\textrm{str}\sim\mathcal{O}(100$~km~s$^{-1})$ and dispersion $\sigma_\textrm{str}~\sim~\mathcal{O}(1$~km~s$^{-1})$. We will assume that the stream comprises some fraction $\rho_\textrm{str}/\rho_a$ of the local dark matter density.

The description of the lab velocity is well known in the context of WIMP direct detection but is not usually considered for axion detection. Since we are reliant on its precise description we will briefly elaborate on its contents. The lab velocity is written,
\begin{equation}
 \textbf{v}_\textrm{lab}(t) = \textbf{v}_0 + \textbf{v}_\textrm{pec} + \textbf{v}_\textrm{rev}(t) + \textbf{v}_\textrm{rot}(t) \, .
\end{equation}
containing respectively, the bulk rotation velocity of the stellar disk (the local standard of rest, LSR), the peculiar velocity of the Sun with respect to the LSR, the orbital speed of the Earth around the Sun and the rotation speed of the Earth about its axis. The latter two velocities are responsible for the annual and diurnal modulations respectively and are known theoretically with effectively perfect precision (see the Appendix of Ref.~\cite{Mayet:2016zxu} for a review of these calculations). In the SHM the velocity of the LSR is usually written as $\textbf{v}_0 = (0,v_0,0)$ in Galactic co-ordinates, with $v_0 = 220$~km~s$^{-1}$. An often quoted value for the peculiar velocity of the Sun from Schoenrich \etal~\cite{Schoenrich:2009bx} is $\textbf{v}_\textrm{pec} = (11.1,12.24,7.25)$~km~s$^{-1}$ with roughly 1~km~s$^{-1}$ sized systematic errors.

\section{Resonance power}\label{sec:experiment}
We model a microwave cavity experiment with a static uniform magnetic field $\textbf{B}_0$ maintained inside a cylindrical cavity of radius $R$ and length $L$, with radial, azimuthal and vertical co-ordinates labelled $(\hat{\textbf{r}},\boldsymbol{\hat{\phi}},\hat{\textbf{z}})$ respectively. The magnetic field is generated by a solenoid with current density in the $\boldsymbol{\hat{\phi}}$-direction. The electric and magnetic fields we write as,
\begin{eqnarray}
  \textbf{E}_0 &=& 0 \\
  \textbf{B}_0 &=& n_L I \Theta (R-r)\hat{\textbf{z}} \, ,
\end{eqnarray}
where $\Theta(r)$ is the Heaviside step function, $I$ is the current and $n_L$ is the number of wire turns in the solenoid per unit length. For convenience we use the magnitude of the magnetic field $B_0 = n_L I$ in the following expressions. 

In the cylindrical cavity design the important cavity mode orientations are the TM$_{0l0}$ modes which have transverse magnetic fields (in the $\boldsymbol{\hat{\phi}}$-direction and hence have associated electric fields in the $\hat{\textbf{z}}$-direction). It is useful to write these induced fields in terms of their Fourier components,
\begin{eqnarray}
\textbf{E}_a &=& E^z_a(r,t)\hat{\textbf{z}} =  \left(\sqrt{T} \int_{-\infty}^{+\infty} \frac{\textrm{d} \omega}{2\pi} E_a(r,\omega) \, e^{-i\omega t} \right) \hat{\textbf{z}} \, , \nonumber \\
\textbf{B}_a &=& B^\phi_a (r,t)\hat{\boldsymbol{\phi}} = \left(\sqrt{T} \int_{-\infty}^{+\infty} \frac{\textrm{d} \omega}{2\pi} B_a(r,\omega) \, e^{-i\omega t} \right) \hat{\boldsymbol{\phi}} \, .\nonumber
\end{eqnarray}
In this case, Amp\`{e}re's law from Maxwell's equations reduces to,
\begin{equation}
 \nabla \times (\textbf{B}_0 + \textbf{B}_a) = \frac{\partial}{\partial t}(\textbf{E}_0 + \textbf{E}_a) - g_{a\gamma\gamma} (\textbf{B}_0 + \textbf{B}_a) \frac{\partial a}{\partial t} \, .
\end{equation}

Solving this equation inside and outside the cavity and matching boundary conditions leads one to a solution for the Fourier components of the axion generated magnetic and electric fields. The solutions are resonances at particular frequencies corresponding to the zeros of a Bessel function (although we will only be interested in the lowest resonance which we label $\omega_0$). Following the derivation of Ref.~\cite{Hong:2014vua}, the axion power is calculated by evaluating the following integral over the volume of the cavity $V$,
\begin{equation}
 P = \frac{\omega_0 U}{Q} = \frac{\omega_0}{Q}\int_V \textrm{d}^3 r \Bigg\langle \frac{\textbf{E}_a^2 + \textbf{B}_a^2}{2}\Bigg\rangle \, ,
\end{equation}
where $U$ is the energy stored in the electric and magnetic fields inside the cavity. This expression introduces the quality factor $Q$ which is a number that quantifies how well the cavity stores energy and depends on the material properties of the cavity wall. Evaluating the above formula with the solution for the Fourier components of the axion electric and magnetic fields (which are expressed in terms of $|\mathcal{A}(\omega)|^2$) one arrives at,
\begin{equation}\label{eq:axionpower}
 P = g_{a\gamma\gamma}^2 B_0^2 V\omega_0 Q^3 \frac{4}{\chi_{0l}^2} \int_{-\infty}^{+\infty} \frac{\textrm{d}\omega}{2\pi} \mathcal{T}(\omega)|\mathcal{A}(\omega)|^2 \, .
\end{equation}
where $\chi_{0l}$ is the $l$-th zero of the 0th Bessel function of the first kind. We have also defined $\mathcal{T}(\omega)$, which is a Lorentzian that describes the loss in power off resonance,
\begin{equation}
 \mathcal{T}(\omega) = \frac{1}{1+4Q^2\left(\frac{\omega}{\omega_0} - 1\right)^2} \, .
\end{equation}

Usually the haloscope power is written in terms of a cavity form factor. For the transverse magnetic field considered here (TM$_{0l0}$) this is written $C_{0l0} = 4/\chi_{0l}^2$\footnote{Other mode orientations, the transverse electric (TE$_{nlm}$) and transverse electromagnetic (TEM$_{nlm}$) modes both have no axial electric field meaning they have negligible form factors.}. We are principally interested in the TM$_{010}$ mode which has $C_{010} = 0.69$. ADMX can tune the TM$_{010}$ mode from roughly 500~MHz to 900~MHz~\cite{Asztalos:2009yp}. In general the electric field of the TM$_{nlm}$ mode can be written~\cite{Jackson},
\begin{equation}
 E_z(r,\phi,z,t) = E(t)J_m{\left(\frac{x_{ml}}{R}r\right)}e^{\pm i m \phi} \cos{\left( \frac{n\pi z}{L} \right)}\, .
\end{equation}
In which, $E(t)$ is the time dependent component of the field, $J_m$ is a Bessel function, $x_{ml}$ is the $l$th root of $J_m(x)=0$, $R$ is the cavity radius and $L$ is the cavity length. Modes with $n\neq0$ and $m\neq0$ have very small form factors.

Our simulation is based upon the calculation of Eq.~(\ref{eq:axionpower}) so for our purposes it would be sufficient to stop here. But in the interest of comparison with previous calculations we will calculate the power on resonance. To do this we simply set $\omega_0 = \omega_a \simeq m_a$ and use a Breit-Wigner approximation for the axion power spectrum with an analogous $Q$-factor $Q_a \sim \omega/\Delta \omega\sim 10^6$ (this permits an analytic evaluation of the integral in Eq.~(\ref{eq:axionpower})). We also introduce the axion density by writing $\langle a^2(t)\rangle = \rho_a/m_a^2$. Resulting ultimately in,
\begin{equation}
 P_a = \hbar^2 c^5 \varepsilon_0 g_{a\gamma\gamma}^2 V B^2 C_{nlm} \frac{\rho_a}{m_a} \, \textrm{min}(Q, Q_a) \,,
\end{equation}
where we have restored the factors of $\hbar$, $c$ and $\varepsilon_0$ for completeness. If the quality factor of the resonant cavity is very high (i.e., the cavity is very good at storing energy and the dissipation is very slow) then the axion conversion power is limited by the spread in axion kinetic energy. The factor $\textrm{min}(Q,Q_a)$ arises from the integral of two Breit-Wigner functions and indicates how the {\it total} power received on resonance is set by the wider of the two power spectra.

Inputting typical values for the experimental parameters we arrive at a total power of the order $10^{-22}$~W as is usually quoted,
\begin{eqnarray}\label{eq:totalpower}
 P_a &=& 6.3 \times 10^{-22} \,\textrm{W} \, \nonumber\\
& \times & \left(\frac{g_{a\gamma\gamma}}{10^{-15} \, \textrm{GeV}^{-1}}\right)^2 \left(\frac{V}{220\, \textrm{l}}\right) \left(\frac{B}{8\, \textrm{T}}\right)^2 \nonumber\\
& \times & \left(\frac{C_{nlm}}{0.69}\right) \left(\frac{\rho_a}{0.3\, \textrm{GeV cm}^{-3}}\right) \nonumber \\
&\times & \left(\frac{3 \, \mu\textrm{eV}}{m_a}\right) \left(\frac{Q}{70,000}\right) \, .
\end{eqnarray}

\section{Mock experiment}\label{sec:analysis}
\begin{figure}
	\includegraphics[width=0.48\textwidth]{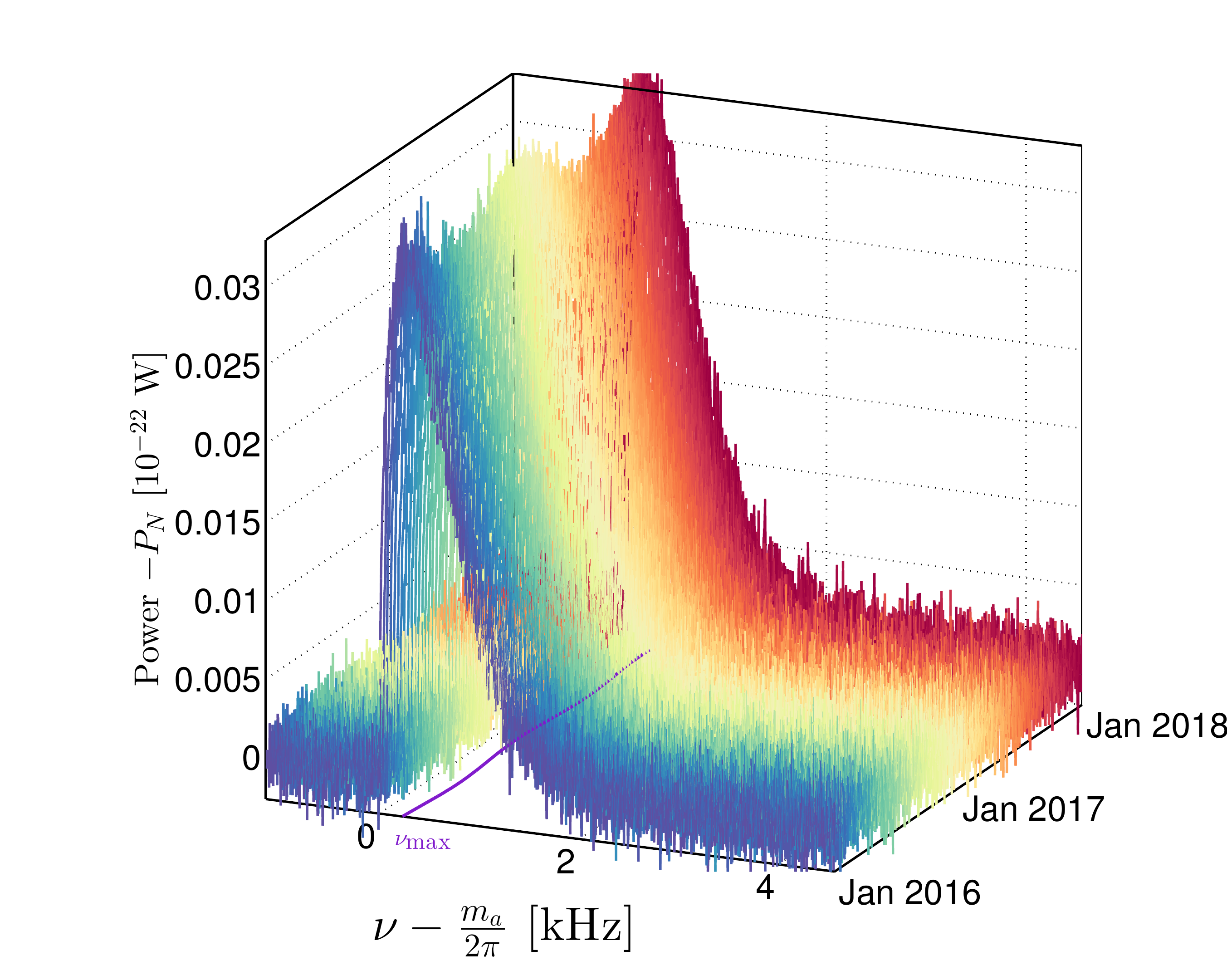}\\
	\includegraphics[width=0.48\textwidth]{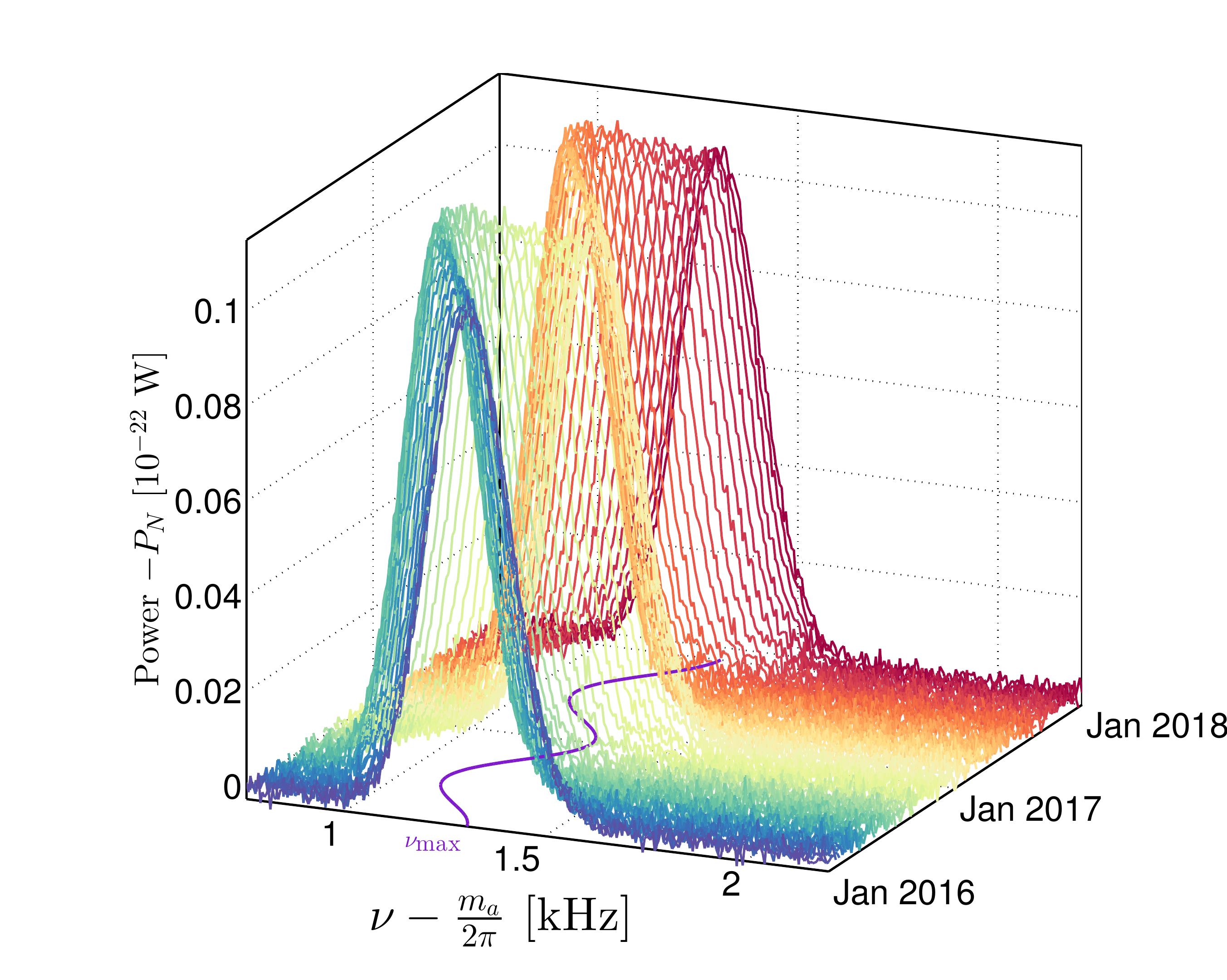}
    \caption{Example simulated power spectra as a function of time. Each line is the average power spectrum observed over a 10 day period. The top panel shows the spectra for a smooth Maxwellian halo and bottom for a pure tidal stream with parameter values displayed in Table~\ref{tab:partable}. The purple line in the frequency-time plane shows the evolution of the frequency at which the power is maximised: $2\pi\nu_\textrm{max} = m_a(1+ v_\textrm{lab}^2/2)$ and $2\pi\nu_\textrm{max} = m_a(1+ |\textbf{v}_\textrm{lab}-\textbf{v}_\textrm{str}|^2/2)$ for the Maxwellian halo and stream respectively.}\label{fig:axionpowerspectrum}
\end{figure}
Our simulation is an approximation of the current ADMX setup. We list a set of benchmark experimental parameters in Table~\ref{tab:partable}. The magnetic field strength, quality factor and noise temperature are roughly in line with what is currently achievable. For calculating the time dependence we also include the latitude and longitude of the experiment.

\begin{table}[t]
  \begin{center}
\begin{ruledtabular}
    \begin{tabular}{l l l}
    
	Axion: 
		& $m_a$ & 3.4671 $\mu$eV \\ 
		& $g_{a\gamma\gamma}$ & 10$^{-15}$ GeV$^{-1}$ \\ \hline 
	Experiment: 
		& $B_0$ & 8 T \\
		& $Q$ & 70,000 \\
		& $V$ & 220 l \\
		& $\Delta \tau$ & 0.2 s \\
		& $\tau$  & 10 days \\
		& $\tau_\textrm{tot}$ & 2 years \\
		& $T_S$ & 4 K \\
		& Latitude & $47.6553^\circ$ \\
		& Longitude & $-122.3035^\circ$ \\ \hline
	Maxwellian halo: 
		& $\rho_a$ & 0.3 GeV cm$^{-3}$ \\ 
		& $\textbf{v}_0$ & $(0,220,0)$ km s$^{-1}$ \\ 
		& $\textbf{v}_\textrm{pec}$ & $(11.1,12.24,7.25)$ km s$^{-1}$ \\  \hline	
	  Stream:
		& $\textbf{v}_\textrm{str}$ & $400 \times (0,0.233,-0.970)$ km s$^{-1}$ \\
		& $\sigma_\textrm{str}$ & 20 km s$^{-1}$ \\
		& $\rho_\textrm{str}$ & 0.05 $\rho_0$ \\
    \end{tabular}
\end{ruledtabular}
  \end{center}
  \caption{Benchmark axion, halo, experimental and stream parameters.}
\label{tab:partable}
\end{table}

In this section we will consider a hypothetical scenario in which the axion has been discovered after a successful low resolution scan over a wider mass range. Once the resonance has been found then an experiment can be performed at a single frequency. The running time of the experiment needs to be long enough to ensure that the signal-to-noise ratio is high but for our purposes also needs to be comprised of long timestream samples to obtain high frequency resolution in the resulting spectrum. 

For now we pick a benchmark set of particle parameters that lie in the QCD axion band: $\nu_a = 842.0$ MHz (= 3.4671 $\mu$eV) and $g_{a\gamma\gamma} = 10^{-15}$~GeV$^{-1}$. This choice evades existing constraints but is easily within the reach of ADMX given a long enough running time at the correct frequency. We use only a single particle benchmark in this study as we are placing the focus on the underlying astrophysical parameters. This is justified however because many of the conclusions are either independent of the choice in mass and coupling (provided the running time and resonant frequency are suitably adjusted) or have dependencies that are simple to explain from the scaling of the axion power. We discuss how one might extend our conclusions to other axion mass and coupling ranges in the Summary Sec.~\ref{sec:summary}.

The sensitivity of a haloscope experiment is limited by the strength of the axion conversion power compared to the noise level. There are two main sources of background noise in resonant cavity experiments: the signal amplifier and the cavity walls. The cavity walls produce thermal blackbody photons or Johnson noise whereas the amplifiers produce electrical noise which depends on the precise technology, however both can be modelled as white noise~\cite{Daw:1998jm,Hotz:2013xaa,Brubaker:2016ktl}. The signal-to-noise ratio for a haloscope experiment of duration $\tau$, is set by the Dicke radiometer equation~\cite{Dicke:1946aa}
\begin{equation}
 \frac{S}{N} = \frac{P_a}{k_B T_S} \sqrt{\frac{\tau}{\Delta \nu_a}} \, ,
\end{equation}
where $\Delta \nu_a$ is the bandwidth of the axion signal and $T_S$ is the noise temperature.

Our mock experiment consists of a long total running time $\tau_{\rm tot}$ which is divided into separate time integrated bins of length $\tau$. Inside a given time bin we calculate a power spectrum which would correspond to the average of $\mathcal{N}$ Fourier transformed timestream samples of duration $\Delta \tau$. The Fourier transform of a given sample is a power spectrum with frequency resolution $\Delta \nu = 1/\Delta \tau$. The noise we simulate as Johnson white noise which has rms power $P_N = k_B T_S \Delta \nu$ inside a given frequency bin with an exponential distribution~\cite{Duffy:2006aa}. The noise power spectrum of the average of $\mathcal{N} = \tau/\Delta \tau$ individual exponential power spectra corresponding to the $\mathcal{N}$ Fourier transformed timestream samples then approaches Gaussian white noise in accordance with the central limit theorem. Hence our simulated noise inside the larger time bin $\tau$ is Gaussian white noise with mean value $P_N$ and standard deviation $P_N/\mathcal{N} = P_N/\sqrt{\tau \Delta \nu}$. The full dataset then consists of a total number of $N_\tau = \tau_{\rm tot}/\tau$ time integrated power spectra each of which consists of the axion power spectrum averaged over the time $\tau$ added to the Gaussian white noise.

The major motivation for a long running time, aside from simply reducing noise, is to utilise the annual modulation due to $\textbf{v}_\textrm{rev}(t)$ which provides a Galactic perspective to the signal. It has previously been shown that the annual modulation signal allows astrophysical parameters to be measured more accurately using WIMP direct detection data, as it breaks degeneracies~\cite{Savage:2006qr}. Below we show that this is also the case for axion searches. We test our simulation by first generating a mock dataset and then attempting to reconstruct the input particle and astrophysical parameters with a maximum likelihood analysis. Two examples of such data are displayed in Fig.~\ref{fig:axionpowerspectrum} corresponding to two halo models, a smooth isotropic Maxwellian distribution and a pure stream (with parameter values listed in Table~\ref{tab:partable}). The annual modulation of the signal is indicated by the purple line labelled $\nu_{\rm max}$.

\begin{figure*}
	\includegraphics[width=0.99\textwidth]{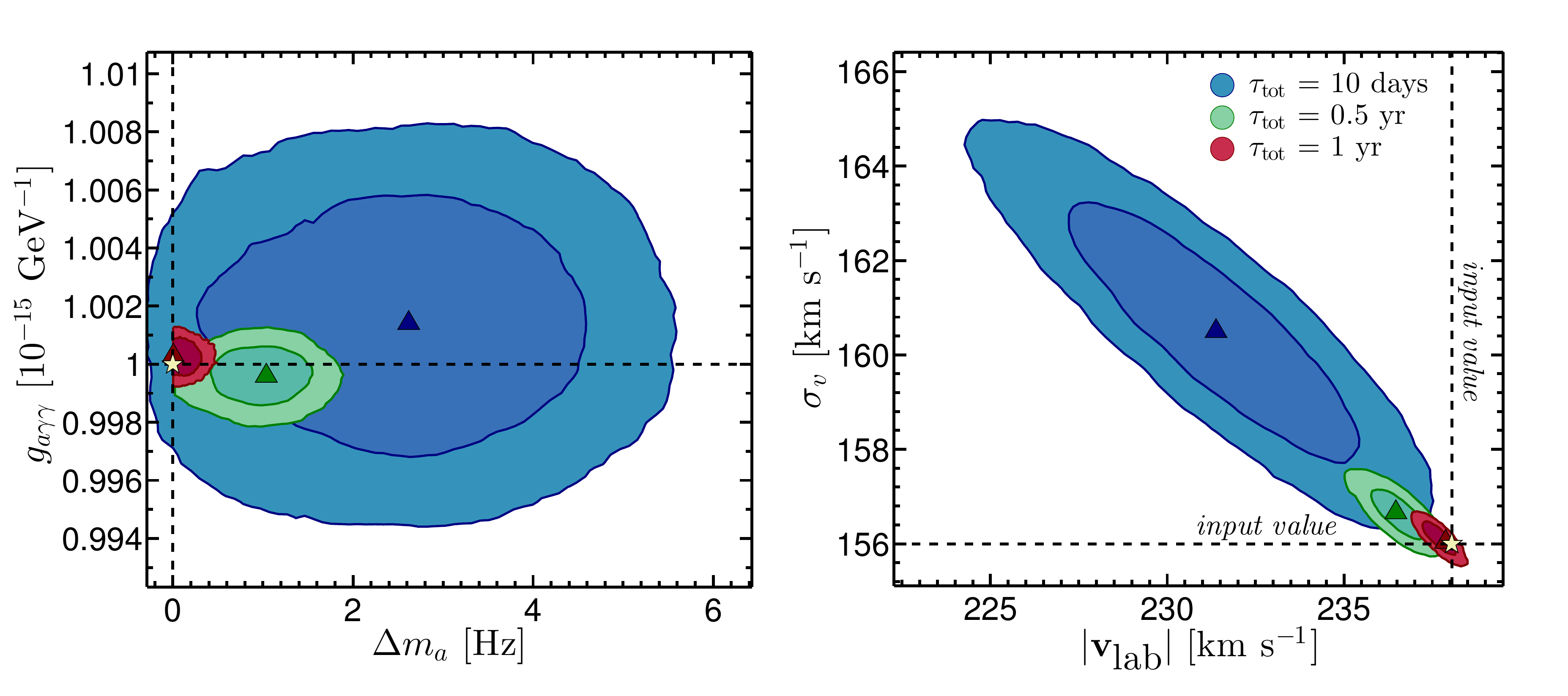}
    \caption{Reconstructed axion mass and coupling as well astrophysical parameters, $v_{\rm lab}$ and $\sigma_v$, for a smooth Maxwellian halo model. We show sets of 68\% and 95\% confidence level contours in the $m_a - g_{a\gamma\gamma}$ and $|\textbf{v}_{\rm lab}|-\sigma_v$ planes (left and right panels respectively). We express the axion mass as $\Delta m_a$ which has the true (input) value subtracted. The blue, green and red sets of contours correspond to the estimates with experiments of different durations: 10 days, half a year and 1 year respectively. The maximum likelihood values are indicated by triangles and the input values for the parameters are indicated by dashed lines and a yellow star.}\label{fig:axion_bench_recon}
\end{figure*}

We base our likelihood on a $\chi^2$ statistic which measures the offset between the observed value of power $P^{ij}_\textrm{obs}$, and the expected power (signal + rms noise) $P^{ij}_a + P_N$ in each bin, where $i$ and $j$ label frequency and time bins respectively,
\begin{equation}
 \chi^2 = \sum_{i = 1}^{N_\nu}\sum_{j = 1}^{N_t} \frac{\left(P^{ij}_\textrm{obs} - P^{ij}_a - P_N\right)^2}{\sigma^2_{N}} \, ,
\end{equation}
where the sums run over $N_\nu = (\nu_\textrm{max}-\nu_\textrm{min})/\Delta \nu$ frequency bins and $N_\tau = \tau_{\rm tot}/\tau$ time bins. The error $\sigma_N$ is given by the suppressed rms noise power $P_N/\sqrt{\tau \Delta \nu}$. We then construct a likelihood based on this statistic. Mathematically the likelihood as a function of a set of parameters given data $\mathcal{D}$ is,
\begin{eqnarray}\label{eq:likelihood}
 \mathcal{L}(m_a,g_{a\gamma\gamma},P_N,\Theta | \mathcal{D}) = e^{-\chi^2/2} \mathcal{L}_\textrm{astro}(\Theta)\, \mathcal{L}_N(P_N) \,,
\end{eqnarray}
where we assume $m_a$, $g_{a\gamma\gamma}$ and $P_N$ are free parameters. We also use the generic $\Theta$ to label a set of astrophysical parameters as we will perform tests with varying numbers of free parameters. We use $\mathcal{L}_\textrm{astro}$ to incorporate the optional uncertainty in the knowledge of an astrophysical parameter (it can be set to unity if no prior knowledge is assumed about a given parameter). The final term $\mathcal{L}_N(P_N)$ parametrises the likelihood of the noise power which can be measured externally (although we set this to unity unless otherwise stated).

Our likelihood analysis consists of first finding the parameter values that maximise the likelihood of Eq.~\ref{eq:likelihood}, we interpret this set of parameters as the best fit points. We then construct 68\% and 95\% confidence regions around these points which are either 1-dimensional intervals when we are only interested in the reconstruction of one parameter or 2-dimensional contours when we are interested in the reconstruction of two parameters and their correlation. We do this by first profiling over all other parameters other than those of interest and then calculate a likelihood ratio between the maximum likelihood for each point $\theta$ in the remaining likelihood function. The likelihood ratio we define as, $\lambda(\theta) = -2(\ln{\mathcal{L}(\theta)}-\mathcal{L}(\hat{\theta}))$, where $\hat{\theta}$ are the maximum likelihood estimators. According to Wilks' theorem~\cite{Wilks:1938dza} this is asymptotically $\chi_k^2$ distributed for $k$ parameters. We then find intervals or contours around the best fit points which enclose regions of parameter values with $\lambda$ less than a certain critical value. The critical value of $\lambda$ is that for which the cumulative distribution of $\chi_k^2$ is the desired confidence level. For example for $k=1$ the 68\% interval encloses values of $\lambda<1$ and the 95\% encloses values of $\lambda<4$.
\section{Reconstructing parameters}\label{sec:reconstruction}
\begin{figure}
	\includegraphics[width=0.49\textwidth]{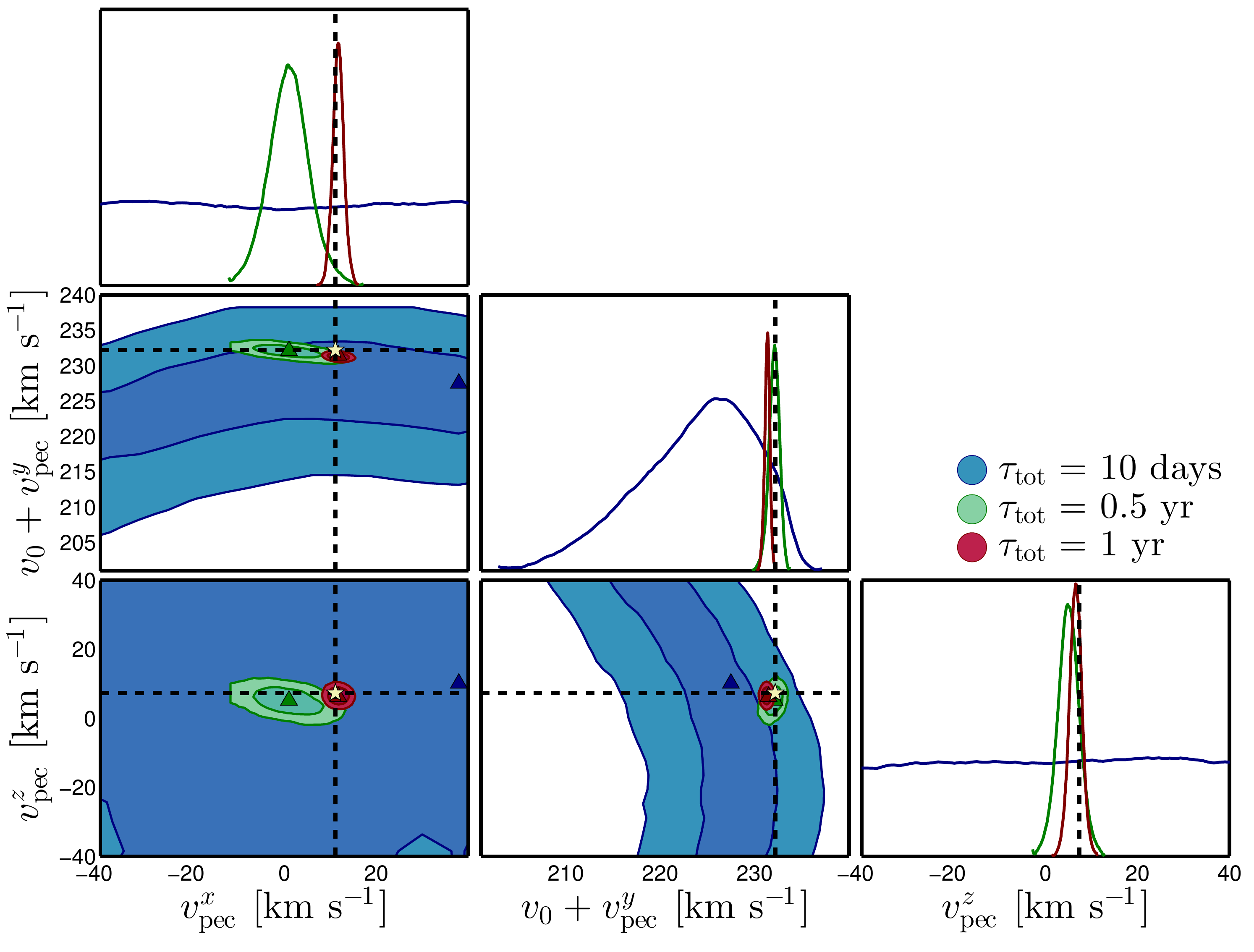}
    \caption{Reconstructed lab velocity components~$(v_{\rm pec}^x,~v_{0}~+v_{\rm pec}^y~,~v_{\rm pec}^z)$  at 68\% and 95\% confidence for three datasets of length 10 days, half a year and 1 year, indicated by blue, green and red sets of contours respectively. The maximum likelihood values are indicated by triangles and the true (input) values are indicated by dashed lines with a yellow star.}\label{fig:vpec_reconstruction}
\end{figure}
\begin{figure}
	\includegraphics[width=0.49\textwidth]{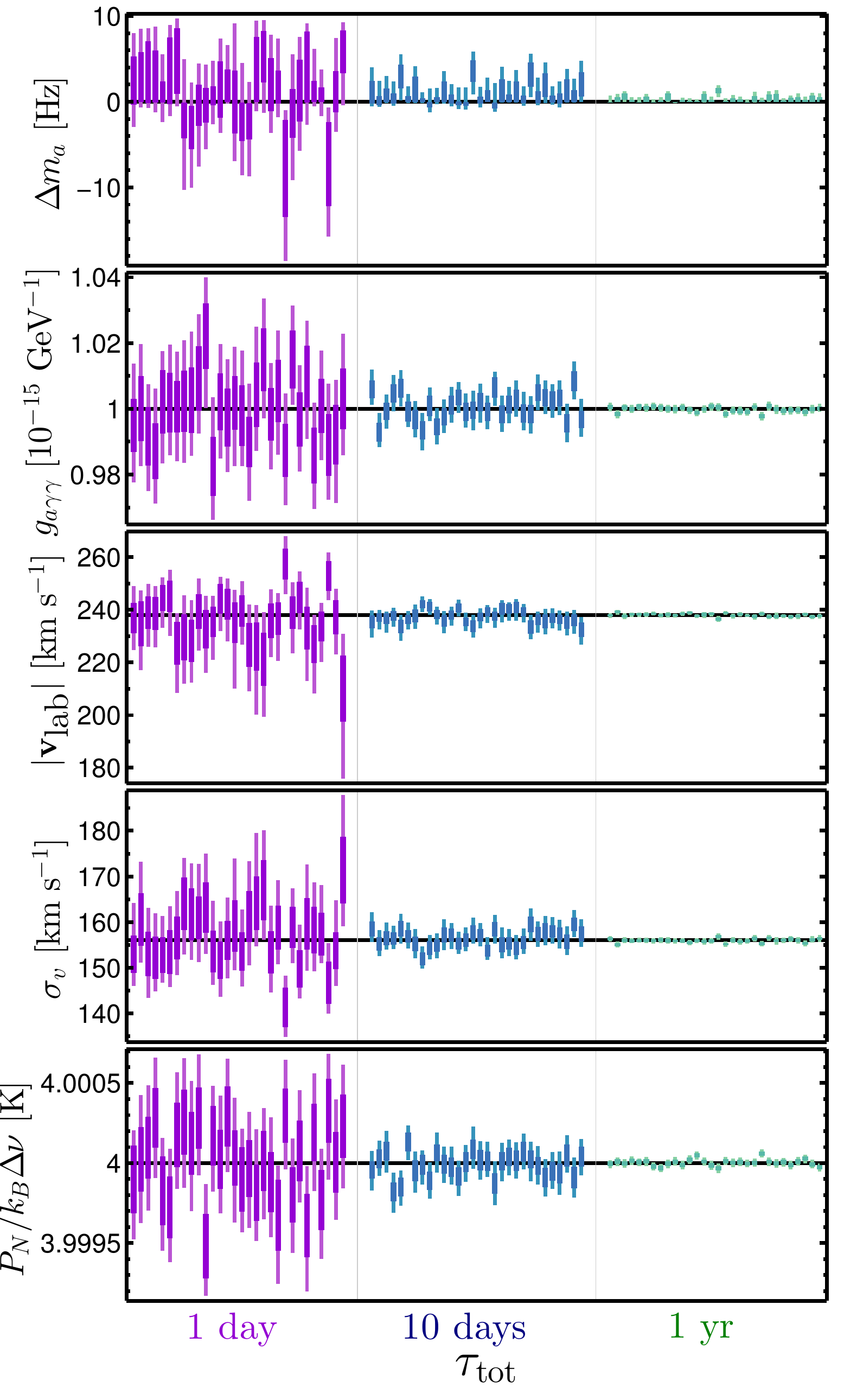}
    \caption{Reconstructed parameters for multiple stochastic data realisations. The 1 and 2 sigma error bars are shown for five parameters, from top to bottom, $m_a$, $g_{a\gamma\gamma}$, $|\textbf{v}_\textrm{lab}|$, $\sigma_v$ and the noise (which we express as $P_N/k_B \Delta \nu$). There are 30 sets of repeated measurements for 3 different experimental durations $\tau_{\rm tot} = $ 1 day, 10 days and 1 year (from left to right).}\label{fig:reconstructionvstau}
\end{figure}
In this section we use the simulation and analysis methodology described in Sec.~\ref{sec:analysis} to attempt to reconstruct sets of input particle and astrophysics parameters. The aim is to quantify how accurately and with what correlations and degeneracies a future ADMX-like haloscope experiment would measure the local axionic dark matter distribution. In the following results we show 1- and 2-dimensional 68\% and 95\% confidence intervals/contours calculated using the profile likelihood, along with best fit parameters values which maximise the likelihood. To explore the likelihood function we use nested sampling algorithms provided by the {\sc MultiNest} package~\cite{Feroz:2007kg,Feroz:2008xx,Feroz:2013hea}, and set a tolerance of $10^{-3}$ and use between $2\times 10^3$ and $10^4$ live points depending on the number of parameters being reconstructed.

In Fig.~\ref{fig:axion_bench_recon} we show the reconstructed axion parameters $m_a$ and $g_{a\gamma\gamma}$ (left) and the astrophysical parameters $v_{\rm lab}$ and $\sigma_v$ (right). We show three sets of contours which correspond to experiments of different durations: 10 days, half a year and 1 year. The 10 day long experiment corresponds to a single time integrated bin of the 0.5 and 1 year long experiments. The annual modulation signal does not play a large role in constraining these parameters, hence the effect of increasing the experiment duration is to shrink the confidence intervals by a factor $\sqrt{\textrm{1 year}/\textrm{10 days}}$.  The axion mass and coupling can be measured to a high level of precision even with only 10 days of data taking, however there is some bias in the best fit values since the dataset consists of a single realisation of stochastic noise. The shapes of the contours are roughly one sided for masses $m>m_a$ due to the fact that the axion power spectrum is only non-zero for $\omega>m_a$. The astrophysical parameters can be measured to a high level of accuracy too. With a 1 year duration the level of precision would reach around the 1~km~s$^{-1}$ level which roughly matches the accuracy of current astronomical observations~\cite{Schoenrich:2009bx}.

With a full annual modulation signal we can also access the 3-dimensional components of $\textbf{v}_\textrm{lab}$. However since $\textbf{v}_0$ and $\textbf{v}_\textrm{pec}$ are summed in the Galactic frame we can only measure directly the $x$ and $z$ components of $\textbf{v}_\textrm{pec}$. The $y$ component (i.e., that which lies along the direction of the rotation of the Milky Way) can only be measured in combination with the LSR speed $v_0$. In Fig.~\ref{fig:vpec_reconstruction} we show the measurement of these parameters for the same three experiment durations of 10 days, 0.5 and 1 year. Since the 10 day duration experiment consists only of a single time integrated bin we have no annual modulation signal and only the reconstruction of the largest component ($v_0+v^y_{\rm pec}$) is possible as this has the greatest influence on the shape of the spectrum. The remaining two components have essentially flat likelihoods as the single time bin spectrum is not sensitive to their values. However for longer durations with modulation in time, the measurement of all three components becomes possible. Even with only half a year of the annual modulation signal we can still make a measurement of the three components of $\textbf{v}_\textrm{lab}$. However, as the signal-to-noise is lower the measurement is biased by particular large fluctuations, which in this example leads to the input values lying outside of the 95\% contour. With a full year of data however a very accurate measurement can be made with 95\% confidence intervals smaller than 5~km~s$^{-1}$ and the true values (indicated by dashed lines and stars) lying within the 95\% interval in all cases.

Finally in Fig.~\ref{fig:reconstructionvstau} we show the 1 and 2 sigma error bars for various parameter measurements as a function of the total experiment duration $\tau_{\rm tot}$. We use three experiment durations from 1 day to 1 year and for each we repeat the experiment 30 times with different randomly generated noise in each to demonstrate the sensitivity to the individual data realisation. As shown in Figs.~\ref{fig:axion_bench_recon} and \ref{fig:vpec_reconstruction} the short duration experiments as well as setting much weaker measurements are also biased by the particular data causing some reconstructions to lie further than 2 sigma away from their input values. In the case of the axion mass we expect one-sided measurements due to the one-sided nature of the power spectrum. This is the case for 10 day and 1 year durations, however for the 1 day duration we see multiple experiments reconstruct a mass smaller than the input mass due to large noise fluctuations in bins slightly below the axion mass. Interestingly for the longer duration experiments the constraint on the axion mass reaches a level smaller than a single frequency bin (5 Hz), this is because the shape of the power spectrum and the annual modulation signal also provide additional information about $m_a$. The size of the error bars for the remaining parameters decrease roughly as $1/\sqrt{\tau_{\rm tot}}$ and for durations long enough to exploit the annual modulation signal we see a significant decrease in the scatter in the reconstructed values over different realisations of the experiment. This means that a future experiment such as this would be able to make fine measurements of the axion particle parameters in conjunction with astrophysical parameters and with no significant biases.

\section{N-body data}\label{sec:nbody}
\begin{figure}
	\includegraphics[width=0.5\textwidth]{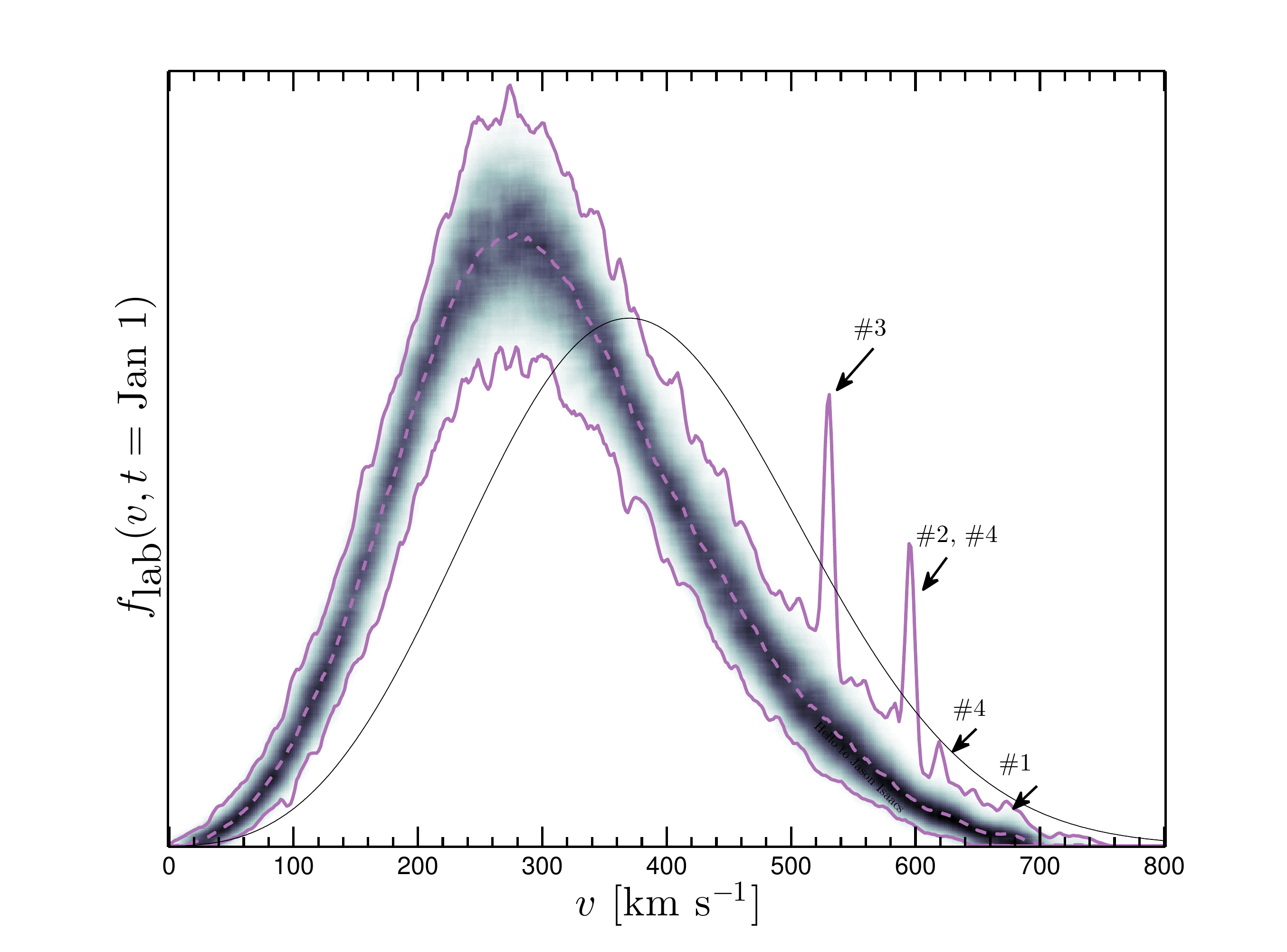}
    \caption{Set of laboratory frame speed distributions of the 200 samples chosen from the VL2 simulation. The shaded regions indicate the range of $f(v)$ values for a given $v$. The solid purple lines indicate the maximum and minimum values of $f(v)$ and the dashed line is the mean distribution over all samples. The black line is the SHM Maxwellian with parameters from Table~\ref{tab:partable}. We label particular samples which contain prominent streams with arrows and the sample number.}\label{fig:vl2speeddist}
\end{figure}
\begin{figure*}
	\includegraphics[width=0.49\textwidth]{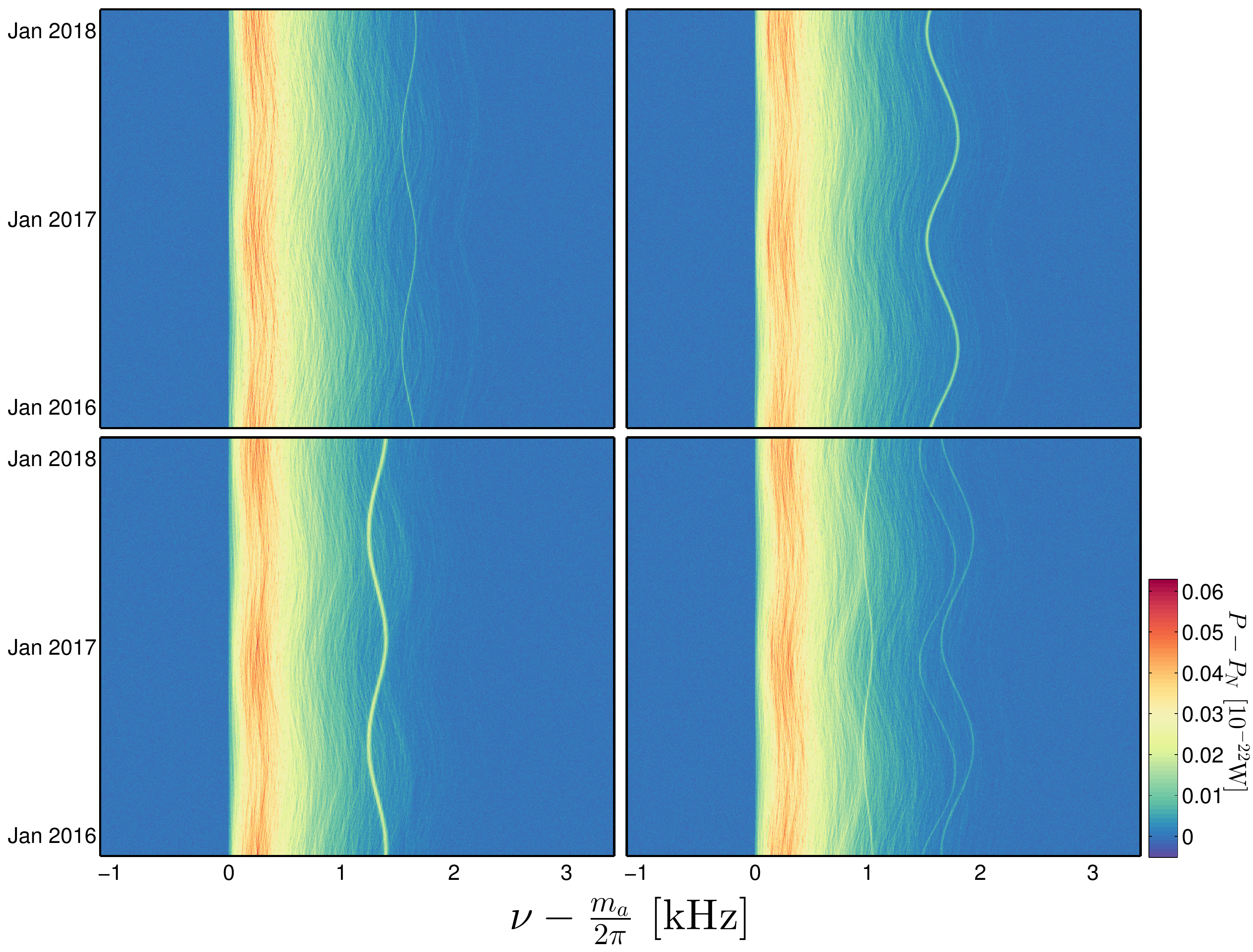}
	\includegraphics[width=0.49\textwidth]{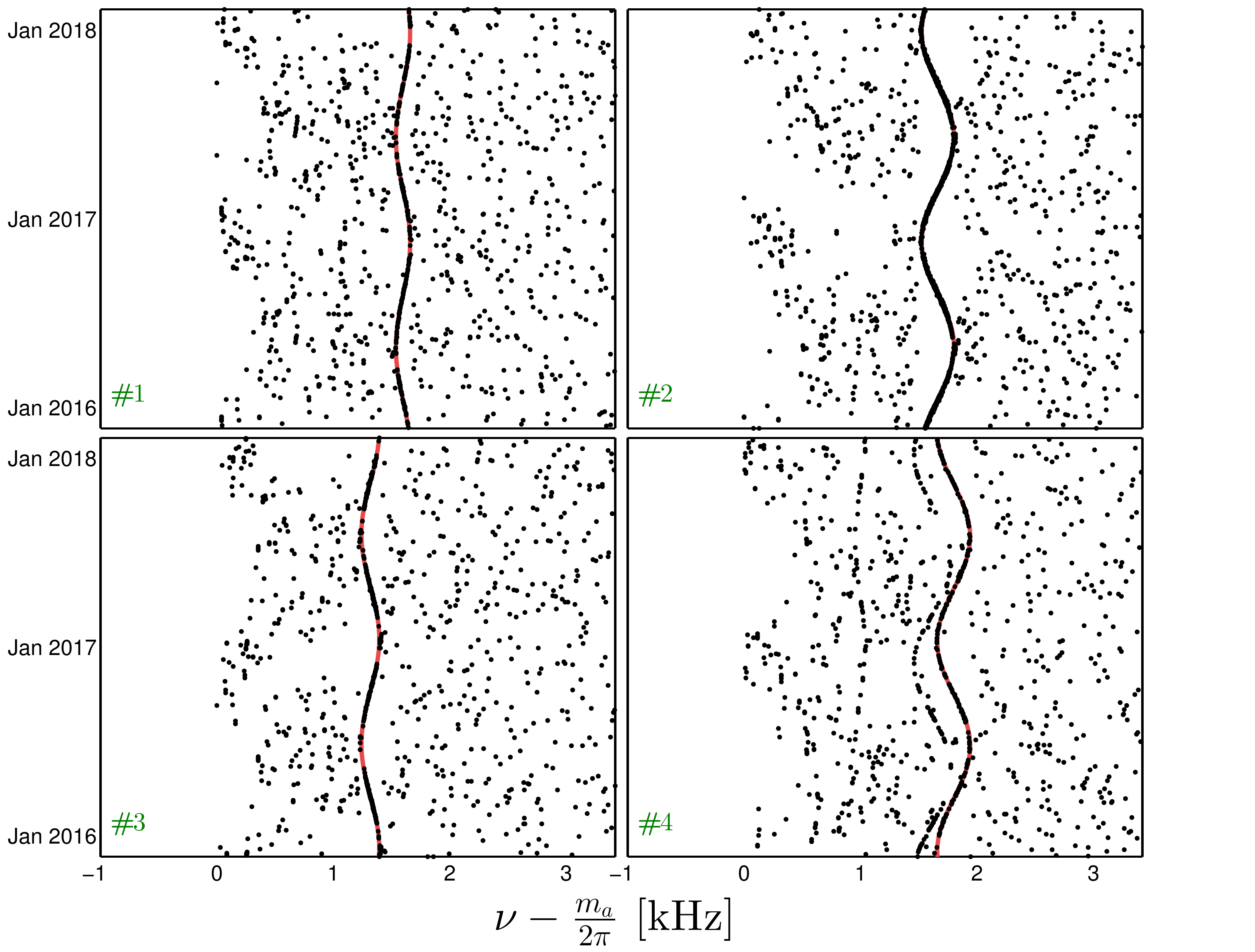}
    \caption{{\bf Left:} Axion conversion power spectra for a selection of four Earth-radius dark matter velocity distributions from the VL2 simulation. In each of the four examples the power spectrum has the amplitude of the noise power ($P_N$) subtracted and is displayed as a function of time (running over 2 years from 2016-2018). The frequency dependence is shown as the difference between the photon frequency and the axion mass. {\bf Right:} The same set of power spectra after performing the various cuts detailed in the text. The remaining points show fluctuations in the axion power spectra after the time independent components have been subtracted. The best fit to Eq.(\ref{eq:sinefit}) is shown as a red line, the power spectrum lying along this best fit line is then extracted to measure the properties of the stream. For clarity in the left hand power spectra we have divided the noise amplitude by 4 so that the substructure is clearer, however the right hand data (used to do the reconstruction) retains the full noise amplitude with $T_S = 4$K.}\label{fig:vl2_powerspectra}
\end{figure*}
\begin{figure*}
	\includegraphics[width=0.99\textwidth]{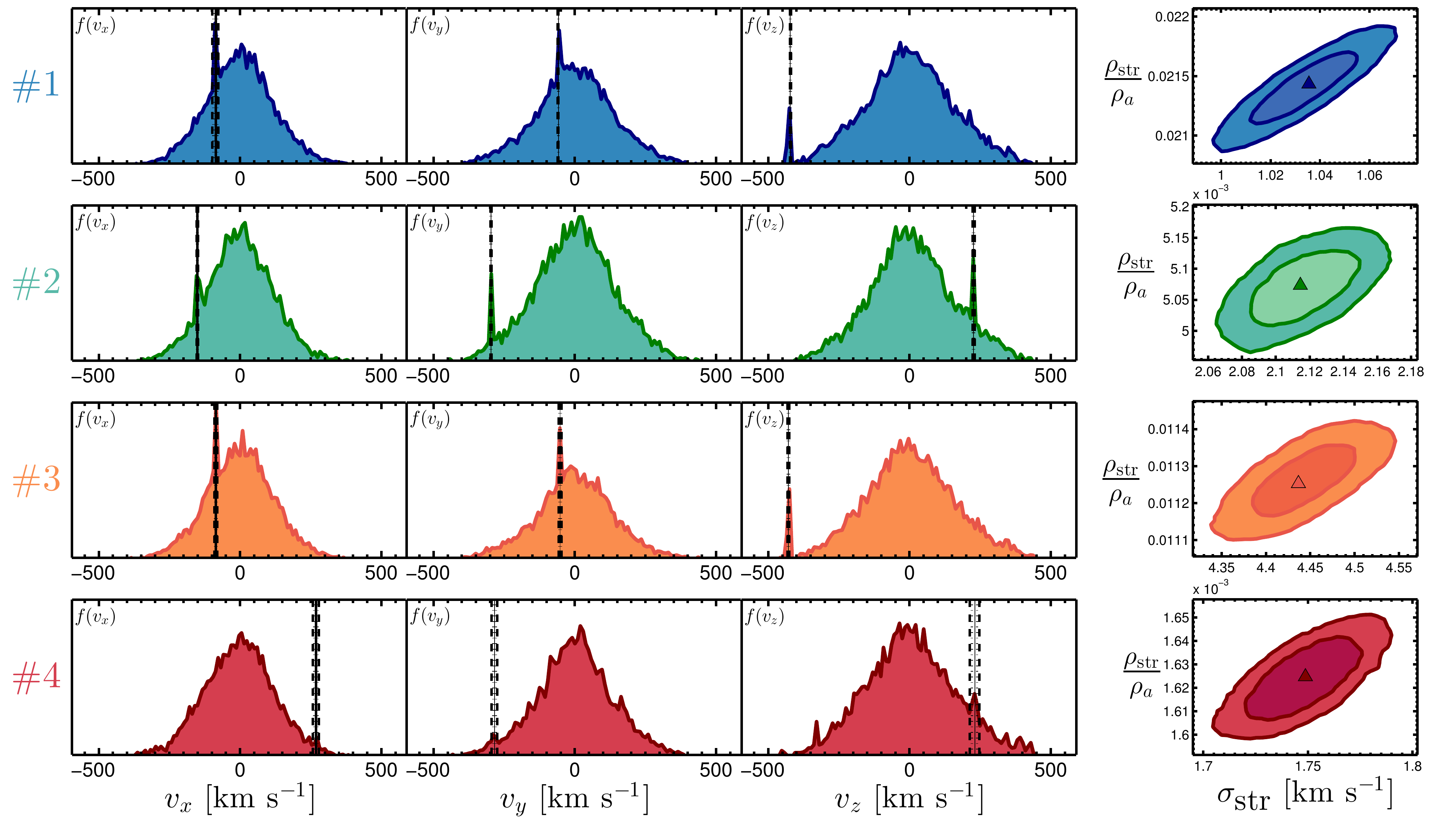}
    \caption{Measurements of stream velocity (vertical black lines) and intervals at the 68\% and 95\% confidence level (dotted and dashed lines respectively) for each of the four sample VL2 velocity distributions. The 1-dimensional speed distributions in each Galactic co-ordinate $(v_x,v_y,v_z)$ correspond to the first three columns. Each row corresponds to the four sample distributions chosen. The final column is the reconstruction in the plane of the stream density fraction and dispersion.}\label{fig:stream_estimate}
\end{figure*}
We can source more realistic examples of dark matter distributions from N-body simulations of Milky Way-like halos. These might more accurately reflect the inhomogeneities and anisotropies that will likely be present in a real dark matter halo. This is of particular interest for a high resolution axion experiment because, as shown in the previous section, it is far more sensitive to astrophysical parameters than standard axion searches and WIMP direct detection. 

We use data from the Via Lactea II (VL2)~\cite{Diemand:2007qr} simulation and select 200 analogue Earth locations at a Galactic radius of 8~kpc and calculate a velocity distribution from all particles contained within 1 kpc spheres centred on each of these locations (we also enforce that no spheres overlap). Although there are more recent hydrodynamic simulations which will better reflect a Milky Way-like dark matter distribution, the VL2 data is sufficient for the illustrative examples we show here and will not change the general conclusions.

We display the range of these 200 velocity distributions in Fig.~\ref{fig:vl2speeddist} with certain samples labelled which contain a significant substructure component. We label these samples from \#1 - \#4. The substructure appearing prominently here are types of tidal stream which are present in real galaxies due to the orbits of satellite galaxies. As these smaller galaxies orbit close to their host halo both the stellar and dark matter components can be tidally stripped leaving a long trail of material around the galaxy which may intersect the main galactic disk. In our own Milky Way there has been evidence from observations and simulations that a particular tidal stream from the Sagittarius dwarf galaxy could pass very close to the location of the Solar System~\cite{Purcell:2012sh}. Tidal streams are of particular interest here as they are very kinematically localised. In particular, streams incoming with velocities at an angle with respect to the motion of the Solar System would give rise to unique annual modulation signals.

We calculate the axion conversion power spectrum in the same way as before but we substitute the analytic $f(\omega)$ with a discretised version calculated by binning particle velocities with a bin size roughly corresponding to the frequency resolution of the experiment. Importantly for each time bin at $t$ we rotate all particle velocities into the laboratory frame with the time dependent Galactic to laboratory transformation detailed in Appendix~\ref{sec:gal2lab}. We must also boost all particle velocities by $\textbf{v} \rightarrow \textbf{v} - \textbf{v}_\textrm{lab}(t)$.

In Fig.~\ref{fig:vl2_powerspectra} we show a selection of four axion conversion power spectra for a range of sample VL2 velocity distributions (the same selection as labelled in Fig.~\ref{fig:vl2speeddist}). The four examples are selected because they contain significant substructure components in the form of streams. These show up in the power spectra as sinusoidally modulating features in time, some examples such as \#2 and \#3 having single dominating streams whereas others such as \#4 possess multiple streams with different amplitudes and phases. 

We can parameterise the frequency dependence of the modulating streams with the function,
\begin{equation}\label{eq:sinefit}
\nu(t) = \nu_1 \sin\left(2\pi\left(\frac{t-t_0}{\textrm{1 year}}\right)\right) + \nu_0 \,.
\end{equation}
In principle the three parameters of this function are related to the three Galactic frame components of the stream velocity, although this will not be a one-to-one mapping. The frequency of the stream modulation $\nu(t)$ is proportional to the quantity $|\textbf{v}_\textrm{str}-\textbf{v}_\textrm{lab}(t)|^2$. 

We can extract substructure components from the data we have presented here by searching for sinusoidally modulating features that have a period of 1 year (whilst also fitting for the function Eq.~(\ref{eq:sinefit})). First we can reduce the data by subtracting the time averaged spectrum and then dividing by the standard deviation of the remaining fluctuations. Next we perform a cut of bins with power fluctuations below a certain level of significance leaving a series of points which if the stream component is large enough will retain the sinusoid modulation. The resulting data points for each example are shown in the left hand panel of Fig.~\ref{fig:vl2_powerspectra}. These data points can then be fit to a model for the stream. We again use the same Maxwellian form for the stream velocity distribution as in Eq.~(\ref{eq:streamfv}) with power spectrum shown in the lower panel of Fig.~\ref{fig:axionpowerspectrum}. Whilst the stream is unlikely to be perfectly described by a Maxwellian, any deviations will be smaller than the error induced by the finite frequency resolution and noise fluctuations. 

A given stream is described by its density $\rho_\textrm{str}$, dispersion $\sigma_{\rm str}$, and three components of velocity $\textbf{v}_\textrm{str}$ making a total of five parameters. Since we have a method for extracting the stream from the data, we can use the data that remains once the stream is removed to fit for the axion, halo and lab velocity parameters and break the degeneracy with the stream parameters. In Fig.~\ref{fig:stream_estimate} we show the reconstructed stream velocities for the four example halo samples displayed in Fig.~\ref{fig:vl2_powerspectra}. Note that in all cases all components of the stream velocity can be reconstructed to high accuracy due to the prominence of the annual modulation signal. This is because the three components of the velocity can all be independently measured with the use of the phase, mean frequency and amplitude of the modulation in Eq.(\ref{eq:sinefit}), although this relationship is nonlinear due to the transformation from the Galactic to the laboratory frame.

Also in Fig.~\ref{fig:stream_estimate} we show the measurement of stream density fraction and dispersion for each sample. Because the density fraction and dispersion are respectively related to the power amplitude and width of the modulating feature, a reconstruction of these parameters is possible in addition to the velocity components. The four samples we have considered here all have relatively large stream contributions which aids the measurement of these parameters. For weaker streams it is likely that longer duration experiments would be required to increase the signal-to-noise. Here the lowest density stream that is detectable with our method is set by the power with respect to the level of noise. Furthermore we have not explored the full stream velocity parameter space with these four examples. It is likely that the accuracy of the reconstruction will be dependent on the direction of the stream with respect to the direction of the lab velocity. Additionally with higher signal-to-noise examples it should also be possible to reconstruct more than one stream (as in sample \#4). We leave these issues however to future work.

\section{Axion miniclusters}\label{sec:miniclusters}
There has been sustained interest in small high density bound structures of axions called miniclusters (see e.g., Refs.~\cite{Hogan:1988mp,Kolb:1993zz,Kolb:1993hw,Kolb:1994fi,Kolb:1995bu,Berezinsky:2013fxa,Tinyakov:2015cgg}). Miniclusters are formed in the early Universe from density perturbations in the axion field. Perturbations which can form miniclusters result from various types of non-linear dynamics involved with axion oscillations such as vacuum misalignment or the decay of axion defects such as strings and domain walls~\cite{Chang:1998tb}. Previous work has predicted the existence of up to $\sim 10^{10}~\textrm{pc}^{-3}$~\cite{Tinyakov:2015cgg} locally if all of the dark matter was in the form of miniclusters, though a direct encounter would occur less than once every $10^5$ years~\cite{Kolb:1994fi}. Through close interactions with stars however axion miniclusters would become tidally disrupted leading to a network of streams wrapping the Milky Way (possibly in addition to a smooth component of the dark matter halo). The miniclusters will pass through the stellar disk many times over the age of the Milky Way ($t_{\rm MW} \sim 12$~Gyr). It has been estimated in Ref.~\cite{Tinyakov:2015cgg} that a small population of disrupted miniclusters would lead to several streams along the path of the Earth through the Galaxy that are large enough to induce an enhancement in the observed total power. The final result of Ref.~\cite{Tinyakov:2015cgg} is a value for the number of expected stream crossings with a density larger than the local smooth halo density $\rho_a$, which is interpreted as an amplification factor. However if the axion minicluster streams are an additional component to the smooth component then the stream density does not need to be larger than the local density to provide an enhancement to the signal. Since the velocity dispersion of the minicluster streams is extremely small compared to the halo ($\sim10^{-4}$~km~s$^{-1}$ $\ll 10^{2}$~km~s$^{-1}$), in a high resolution axion experiment all of the minicluster stream axions would convert to photons in a small number of frequency bins. Hence for a minicluster stream to be observable we simply need the total power from the stream to be larger than the power over a few bins.

Individual miniclusters are parameterised by the density contrast in the axion field, $\Phi = \delta \rho_a / \rho_a$ which is a number typically of order unity. The distribution of values of $\Phi$ found from the simulations of Ref.~\cite{Kolb:1995bu} appears to follow a function similar to $f_\Phi(\Phi) \sim \Phi^{0.75} e^{-\Phi}$ which we will use as an approximation. The mass of a minicluster is set by the total mass of axions inside the Hubble radius around the time when axion oscillations begin, $M_1 \sim 10^{-12}\, M_\odot$ (which is allowed by lensing bounds~\cite{Zurek:2006sy}). Ref.~\cite{Kolb:1995bu} states that the distribution of minicluster masses is concentrated tightly around a large fraction of this mass.

Solving for the collapse of a spherical region with density contrast $\Phi$ and evolving through cosmic time to the present day gives a range of minicluster densities, 
\begin{equation}
\rho_{\rm mc}(\Phi) \simeq 7 \times 10^6 \, \Phi^3 (1+\Phi) \, \textrm{GeV cm}^{-3} \, .
\end{equation}
We assume that the miniclusters are spherically symmetric with central density $\rho_{\rm mc}(\Phi)$ and radius $R_{\rm mc}(\Phi,M)$. We assume a Maxwellian distribution for the speeds of axions inside a minicluster with a dispersion set by the virial velocity $\sigma_{\rm mc}(\Phi,M) = \sqrt{G M/R_{\rm mc}(\Phi,M)}$. From Ref.~\cite{Zurek:2006sy} we assume the miniclusters have a power law density profile with $\rho \propto r^{-9/4}$ for $r<R_\textrm{mc}$ but enforce the gradient to turn towards 0 at $r=0$ to give a central density of $\rho_{\rm mc}(\Phi)$. 

The number of streams expected at the Solar radius results from evolving the initial distribution of axion miniclusters through the age of the Galaxy to today. Each time the minicluster crosses the stellar disk there is a probability that it will encounter a star close enough to become disrupted. Following previous calculations of this type~\cite{Goerdt:2006hp,Schneider:2010jr}, Ref.~\cite{Tinyakov:2015cgg} gives the probability of a particular minicluster being disrupted,
\begin{equation}\label{eq:disruptprob}
P(\Phi) = 8\pi n S_{\perp} \frac{G R_{\rm mc}(\Phi,M)}{v \sigma_{\rm mc}(\Phi,M)} \,,
\end{equation}
where here $v$ is the orbital speed of the minicluster, and $n$ the number of crossings of the stellar disk the minicluster undergoes. This calculation has already averaged over an isotropic distribution of minicluster trajectories and has been written in terms of the stellar contribution to column density in the direction perpendicular to the disk, $S_{\perp} = 35 \,\textrm{M}_\odot\, \textrm{pc}^{-2}$~\cite{Kuijken:1989hu}. Given this, we can just use miniclusters with circular orbits intersecting the Solar position ($r_\odot$) to evaluate the number of crossings over the age of the Galaxy ($t_{\rm MW}$) to be roughly $n \sim 2 \, t_{\rm MW}/t_{\rm orb} \sim v \,t_{\rm MW} / \pi r_\odot \sim 100$. Note that the dependence on $v$ drops out of Eq.~(\ref{eq:disruptprob}). This is because although faster miniclusters cross the stellar disk more frequently ($\propto v$), they are also less likely to encounter a star during a given crossing ($\propto 1/v$). We also note that $P(\Phi)$ has no dependence on $M$ since $R_{\rm mc}(\Phi,M)$ and $\sigma_{\rm mc}(\Phi,M)$ are both proportional to $M^{-1/3}$.
 
A stream can be specified alone by four parameters: the density contrast $\Phi$ and mass $M$ of the original minicluster, the age of the stream $t$, and the orbital velocity of the minicluster/stream, $\textbf{v}$. All other parameters can be derived (we indicate dependence on each by parentheses). Once a minicluster is disrupted by a star it will begin to leave a trail of axions along its orbit, the length of which will stretch linearly with time as the cluster orbits the Galaxy $L \sim \sigma_{\rm mc} t$. Assuming the stream retains the original radius of the minicluster and is simply deformed from a sphere of radius $R_{\rm mc}$ into a cylinder of length $L$, the density of the axions for a minicluster stream of age $t$ is, 
\begin{equation}\label{eq:mcstreamdensity}
\rho_{\rm str}(\Phi,M,t) = \rho_{\rm mc}(\Phi) \frac{\frac{4}{3}R_{\rm mc}(\Phi,M)}{\sigma_{\rm mc}(\Phi,M) t} \, ,
\end{equation}

Reference~\cite{Tinyakov:2015cgg} calculates the number of expected stream crossings in a 20 year period for two values for the age of the Galaxy and two masses. We extrapolate the final result of this work down to stream densities of $\rho_a/N_\nu \sim 0.001 \rho_a$ as this is around the lowest density stream that would be observable in this case. We estimate that if this extrapolation of Ref.~\cite{Tinyakov:2015cgg} is valid then, for $t_{\rm MW} = 12$~Gyr and $M = 10^{-12} M_\odot$, there could be up to $N_{\textrm{str-x}} \sim 100$ stream crossings in a 20 year period (although the precise number is not important for the illustrative example we present here). 

We simulate the signal for $N_\textrm{str-x}$ minicluster stream crossings by selecting samples from the parameter space $\{\Phi, \textbf{v}, \rho_\textrm{str}\}$. First we select values for $\Phi$ from the distribution $P(\Phi)f_\Phi(\Phi)$. We then select $\textbf{v}$ from an isotropic Maxwell-Boltzmann distribution. Finally we draw a value of $\rho_\textrm{str}$ such that the number of stream crossings with $\rho_\textrm{str}>\rho_a$ follows the function presented in Fig. 2 of Ref.~\cite{Tinyakov:2015cgg}. The length of time taken to cross the stream is then approximately,
\begin{equation}\label{eq:streamcrossingtime}
\tau_\textrm{str-x}(\Phi,M,\textbf{v}) = \frac{2 R_{\rm mc}(\Phi,M)}{v_\textrm{lab}\sqrt{1-\left(\frac{\textbf{v}_\textrm{lab}\cdot\textbf{v}}{v_\textrm{lab}v}\right)^2}} \, .
\end{equation}
which is derived from the distance travelled through the stream, $2R_\textrm{mc}/\sin\theta$, where $\theta$ is the angle between the stream and the path of the Earth. We distribute each of these crossings uniformly over the running time of the experiment. The power spectrum observed during a crossing is enhanced with an additional Maxwellian component (as with the streams the previous section) with relative velocity $\textbf{v}_\textrm{lab} - \textbf{v}$ and dispersion $\sigma_{\rm mc}$. Also in a given time bin the minicluster stream signal will gain an additional spread in frequency from the change in $\textbf{v}_\textrm{lab}(t)$ over the duration of the bin.

To deal with Eq.~(\ref{eq:streamcrossingtime}) diverging for stream directions that align with the path of Earth we remove all streams which orbit with $\tan \theta<\frac{1}{2}z_\textrm{disk}/r_\odot$ relative to the plane of the stellar disk, where $z_\textrm{disk}\sim 0.3\,{\rm kpc}$ is the width of the stellar disk. This is a safe approximation as this is only a small fraction of the streams, and miniclusters that orbit in the plane of the stellar disk will become disrupted much earlier than those orbiting at a large angle and the streams will hence have much lower present day densities.

\begin{figure}
	\includegraphics[width=0.5\textwidth]{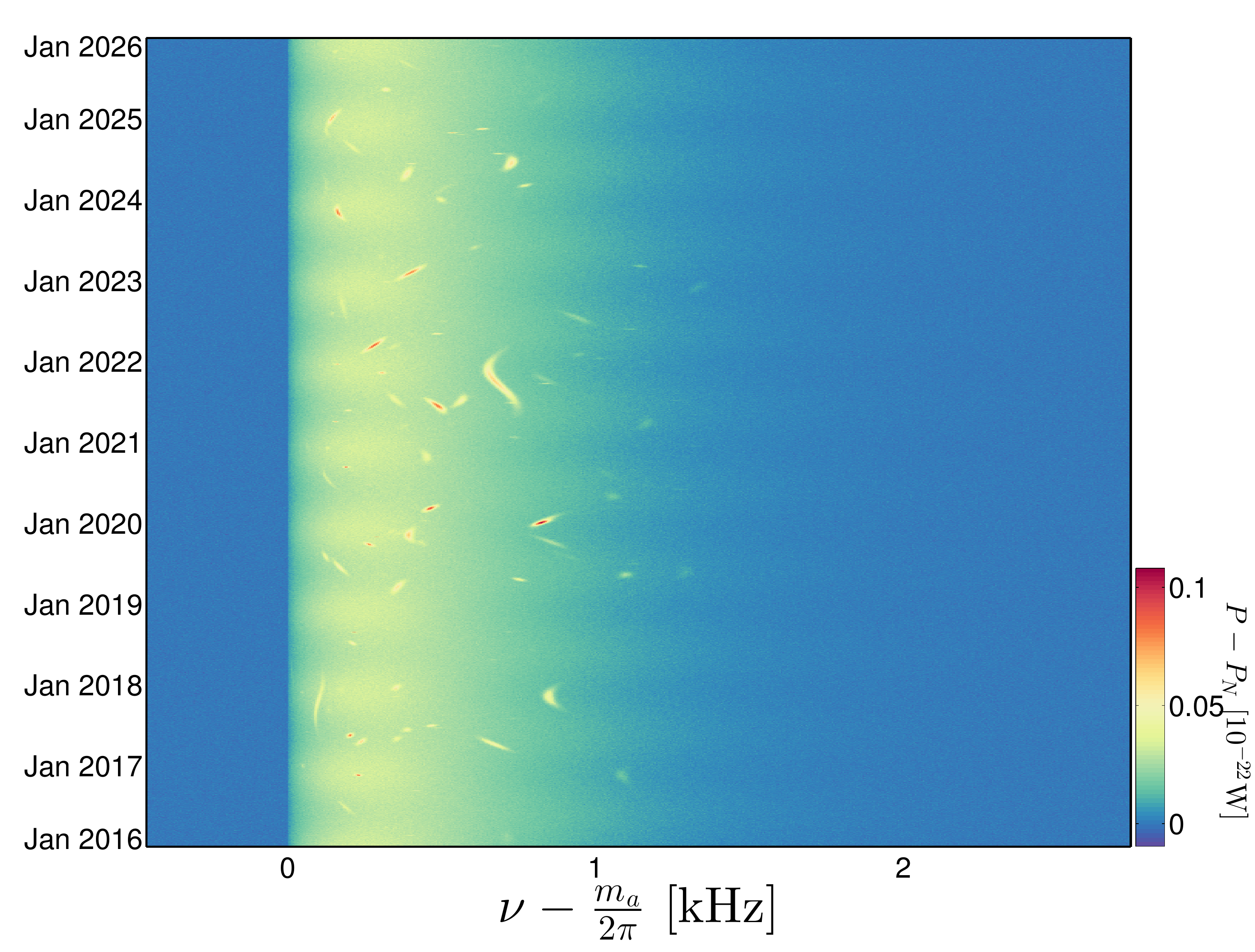}
    \caption{Simulated power spectra observed over a 10 year period for a halo model consisting of a smooth population of axion dark matter with an additional component from a network of tidal streams stripped from orbiting miniclusters. The abundance is based on the calculation of Ref.~\cite{Tinyakov:2015cgg}. The signal from minicluster streams appears as short-lived enhancements which are modulated in frequency due to the orbit of the Earth. The power spectra are displayed as a function of time from Jan 2016 to Jan 2021 and frequency shifted by the axion mass.}\label{fig:miniclusterpowerspectrum}
\end{figure}

In Fig.~\ref{fig:miniclusterpowerspectrum} we display a simulated power spectrum observed over a total period of 10 years for a halo consisting of a smooth population of axions and a network of tidal streams from miniclusters. The streams appear as peaks in the power spectrum over a very narrow range of frequencies (as in Sec.~\ref{sec:nbody}) but here since minicluster radii are on the scale of $10^7$~km they are short-lived enhancements compared with usual tidal streams which extend over volumes larger than the scale probed by the Galactic orbit of the Solar system. 

The total power measured in the form of these short-lived enhancements would provide an estimate of the fraction of local axion dark matter contained in minicluster streams from which the abundance of miniclusters could be inferred. We emphasise however that a detailed theoretical treatment of the disruption of a population of miniclusters is still needed in order to fully explore the prospects for their detection. Our example here shows that even if miniclusters comprise only a very small contribution to the local axion density, they appear much more prominently in a high resolution experiment. In principle one could make use of the methods described in Secs.~\ref{sec:reconstruction} and~\ref{sec:nbody} to extract information about individual streams such as their radius, age and Galactic frame velocity as well as place constraints on the minicluster population such as their mass spectrum and abundance. This would require isolating the signal from miniclusters from both the noise and the background axion power spectrum. A possible strategy could be to use the observations during periods without any minicluster enhancement to make accurate measurements of the underlying parameters (as in Sec.~\ref{sec:reconstruction}) to then subtract the background spectrum thus isolating the stream crossing events. 

A further complication that we have not discussed in this work is the presence of any short-lived environmental peaks which may appear in real resonant cavity experiments and could mimic a positive axion signal. These would usually be dealt with by performing a repeat experiment in the frequency range at which the peak was observed. However in the case of minicluster streams which are themselves short-lived this check would not necessarily be successful if the timescales for the environmental peak and the stream crossing were comparable. However a careful treatment of the frequency modulation of the peak over time may in some cases be enough to distinguish a Galactic signal from a lab-based one. We leave a more detailed study of axion minicluster streams and implications for experiments to a future work.

\section{Summary}\label{sec:summary}
We have performed a simulation of a hypothetical high resolution ADMX-like experiment following a successful detection of an axion dark matter signal. Our focus here has been on extracting astrophysical information and performing axion astronomy. Our main conclusions are as follows:
\begin{itemize}
\item The measurement of the axion-photon conversion power spectrum enables the accurate reconstruction of both axion particle parameters in conjunction with the underlying astrophysical parameters.
\item With the use of the annual modulation signal one can make accurate measurements of the components of the Solar peculiar velocity. With an experimental duration longer than a year the accuracy can reach below 1 km s$^{-1}$, which would improve upon the measurement from local astronomical observations.
\item Substructure such as tidal streams appearing in simulations of Milky Way-like halos show up prominently in the resolved axion power spectrum and can hence be measured to levels of sensitivity not possible in the direct detection of WIMPs. The annual modulation signal plays an important role here too as the precise shape of the modulating stream allows the reconstruction of its properties: the Galactic frame velocity, density and dispersion. This in principle would allow axion haloscopes to trace the formation and accretion history of the Milky Way.
\item  We have simulated an approximation to the expected signal from a population of streams from disrupted axion miniclusters. We have extrapolated a result for the calculation of the expected number of stream crossings from Ref.~\cite{Tinyakov:2015cgg}. In an experiment that resolves the axion spectrum the signal from minicluster streams would appear much more prominently in the data and could be isolated to place constraints on their mass spectrum or abundance.
\end{itemize}

The issues we have discussed here are relatively unstudied in the context of axion detection. Hence there are a number of areas in which this study might be extended. We have shown that measuring the axion power spectrum allows accurate reconstruction of underlying parameters and although we have only considered simple models here, in principle the same should be true of other models for the dark matter velocity distribution such as those containing anisotropy parameters or additional substructure such as debris flows~\cite{Kuhlen:2012fz}, dark disks~\cite{Schaller:2016uot} and caustic rings~\cite{Duffy:2008dk}. What remains to be seen however is the extent to which the correct selection of a particular model is possible with data of this kind. This is an important consideration for WIMP direct detection experiments with very low statistics, multiple competing experiments and degeneracy between assumptions about the underlying velocity distribution. These issues have given rise to a number of approaches for making astrophysics independent limits and measurements~\cite{Frandsen:2011gi,Fox:2010bu,Gondolo:2012rs,DelNobile:2013cta,Fox:2014kua,Feldstein:2014gza,Kahlhoefer:2016eds,Gelmini:2016pei} and developing general parameterisations for the velocity distribution~\cite{Peter:2011eu,Kavanagh:2013wba,Kavanagh:2013eya,Kavanagh:2016xfi}. In the case of axions however, because the power spectrum could be measured to an arbitrary level of precision given sufficient duration it may not be necessary to develop any such astrophysics independent methods, however this would require a separate investigation. 

We have used only one axion benchmark mass and coupling, since our focus is on measuring the underlying astrophysical parameters. However our conclusions can be simply extended to other values by considering Eq.(\ref{eq:totalpower}). Since the total axion power is proportional to $g_{a\gamma\gamma}^2$ one can extend any of the reconstructions to smaller couplings by scaling up the experiment duration, $\tau_{\rm tot}$, by the same factor squared. Although it should be noted that haloscopes can reach smaller couplings by both reducing noise as well as simply extending the duration of the experiment and both of these approaches are necessary for improving constraints on the axion coupling. Since the total power is proportional to $m_a^{-1}$, our conclusions still hold for smaller (larger) values of the axion mass if $\tau_{\rm tot}$ is increased (decreased) by the same factor. The reverse argument goes for values of the local density since the power is linearly proportional to $\rho_a$. However we must take care in extending these results to axion masses much larger or smaller than $\mathcal{O}(\mu$eV) since a given experimental design is only able to probe masses in a small range. There are several reasons for this. First, it is the frequency range of the experiment that dictates the range of axion masses that can be probed. ADMX is suited to masses $<10\, \mu$eV and has currently set constraints between $1.9\,\mu{\rm eV} < m_a < 3.69\,\mu{\rm eV}$~\cite{Asztalos:2009yp,Hoskins:2011iv}. Larger masses require adjustments to the cavity and amplification technology~\cite{Slocum:2014gwa,Baker:2011na}. The Yale Wright Laboratory experiment of Refs.~\cite{Brubaker:2016ktl,Kenany:2016tta} for example operates between 5 - 25 GHz (corresponding to 20 - 100 $\mu$eV) and is the first to set limits for $m_a>20 \mu$eV over a 100 MHz range. A number of experimental challenges are present in designing experiments for different mass windows. For higher resonant frequencies the effective volume of the cavity falls off quickly as $\nu^{-3}$ meaning the cavity geometry must be revised to preserve form factors and thus maintain the sensitivity of the experiment. There are also limitations on the frequency ranges for which the SQUID amplification technology is useful meaning new techniques must be developed such as Josephson parametric amplifiers~\cite{Kenany:2016tta} for the GHz range. For masses towards 40~-~400 $\mu$eV the dielectric disk setup of MADMAX~\cite{TheMADMAXWorkingGroup:2016hpc,Millar:2016cjp} has been designed and avoids the restriction placed on resonators brought about by the dependence on the cavity volume. Smaller masses $10^{-(6-9)}$~eV may also be accessible with nuclear magnetic resonance-based experiments such as CASPEr~\cite{Budker:2013hfa,Graham:2013gfa} or alternative designs with resonant and broadband readout circuits~\cite{Kahn:2016aff}, and LC circuits~\cite{Sikivie:2013laa}.

Ultimately the prospects for axion astronomy will depend on the success of one of the aforementioned search strategies, at which point the development of the optimum technology to measure dark matter axion-photon conversion can begin. In addition to the annual modulation signal, which we have shown to be powerful for making more accurate measurements of some astrophysical parameters, it may also be beneficial to search for possible direction dependent methods (e.g., Refs.~\cite{Irastorza:2012jq,Horns:2012jf,Jaeckel:2015kea}) as the angular signature of a dark matter signal has been shown to encode much astrophysical information in the context of WIMPs~\cite{Lee:2012pf,OHare:2014nxd}. However in any of these possible scenarios the methods developed in this study will be a valuable step in progressing toward axion astronomy.

\acknowledgments
The authors are grateful for useful correspondence with Igor Irastorza and Benjamin Brubaker. C.A.J.O. is supported by a United Kingdom Science and Technology Facilities Council (STFC) studentship. A.M.G. acknowledges support from STFC Grant No. ST/L000393/1.

\appendix
\section{Galactic to Laboratory transformation}\label{sec:gal2lab}

Here we briefly summarise the co-ordinate transformation used to rotate particle velocities from the Galactic system into the rest frame of the laboratory. 

The Galactic co-ordinate system $(\hat{\textbf{x}}_g,\hat{\textbf{y}}_g,\hat{\textbf{z}}_g)$ is defined such that $\hat{\textbf{x}}_g$ points towards the Galactic center, $\hat{\textbf{y}}_g$ points in the plane of the Galaxy towards the direction of Galactic rotation and $\hat{\textbf{z}}_g$ points towards the Galactic North pole. We define the Laboratory co-ordinate system $(\hat{\textbf{x}}_\textrm{lab},\hat{\textbf{y}}_\textrm{lab},\hat{\textbf{z}}_\textrm{lab})$ which point towards the North, West and zenith respectively. To move between these co-ordinate systems we also need an intermediate step in the geocentric equatorial frame $(\hat{\textbf{x}}_e,\hat{\textbf{y}}_e,\hat{\textbf{z}}_e)$, where $\hat{\textbf{x}}_e$ and $\hat{\textbf{y}}_e$ point towards the celestial equator with right ascensions of 0 and 90$^\circ$ respectively and $\hat{\textbf{z}}_e$ points to the celestial north pole.

We transform vectors (e.g., VL2 particle velocities) from the Galactic to the laboratory frame with the following transformation,
\begin{equation}
 \begin{pmatrix}\hat{\textbf{x}}_\textrm{lab}\\\hat{\textbf{y}}_\textrm{lab}\\\hat{\textbf{z}}_\textrm{lab}\end{pmatrix} = \textbf{A}_{e\rightarrow\textrm{lab}}\left(\textbf{A}_{g\rightarrow e} \begin{pmatrix}\hat{\textbf{x}}_g\\\hat{\textbf{y}}_g\\\hat{\textbf{z}}_g\end{pmatrix} \right) \, ,
\end{equation}
where the transformation from the Galactic to equatorial system is encoded in the matrix~\cite{Bozorgnia:2011vc},
\begin{eqnarray}
 &\textbf{A}_{g\rightarrow e} =& \\
&\begin{pmatrix}
-0.05487556 & +0.49410943 & -0.86766615 \\
-0.87343709 & -0.44482963 & -0.19807637 \\
-0.48383502 & +0.74698225 & +0.45598378
\end{pmatrix}  \, , & \nonumber
\end{eqnarray}
and the transformation to the laboratory frame is given by,
\begin{eqnarray}
 &\textbf{A}_{g\rightarrow \textrm{lab}}  = &\\
&\begin{pmatrix}
 -\sin(\lambda_\textrm{lab})\cos(t^\circ_\textrm{lab}) & -\sin(\lambda_\textrm{lab})\sin(t^\circ_\textrm{lab}) & \cos(\lambda_\textrm{lab}) \\
 \sin(t^\circ_\textrm{lab}) & -\cos(t^\circ_\textrm{lab}) & 0\\
 \cos(\lambda_\textrm{lab})\cos(t^\circ_\textrm{lab}) & \cos(\lambda_\textrm{lab})\sin(t^\circ_\textrm{lab}) & \sin(\lambda_\textrm{lab})
\end{pmatrix} &\nonumber \, .
\end{eqnarray}
In which we have used $\lambda_\textrm{lab}$ for the Earth latitude of the laboratory. We use $t^\circ_\textrm{lab}$ for the local apparent sidereal time expressed in degrees which is related to the Julian day (JD) by the following,
\begin{eqnarray}
 t^\circ_\textrm{lab} &=& \phi_\textrm{lab} + \bigg[101.0308 \nonumber \\
		& +& 36000.770\left(\frac{\floor{\textrm{JD} - 2400000.5}-55197.5}{36525.0}\right) \nonumber \\
		& +& 15.04107 \, \textrm{UT}\bigg] \, ,
\end{eqnarray}
where $\phi_\textrm{lab}$ is the longitude of the laboratory location. We also must convert the Julian day to Universal Time (UT) using,
\begin{equation}
 \textrm{UT} = 24\,\bigg(\textrm{JD}+0.5-\floor{JD+0.5}\bigg) \, .
\end{equation}

\bibliographystyle{apsrev4-1}
\bibliography{axions.bib}

\end{document}